\documentclass{article}

\usepackage[preprint,nonatbib]{neurips_2026}
\usepackage{custom}

\title{A simple model of co-emergence of grid and place fields}

\author{%
Zhaoze Wang\textsuperscript{1} \\
\texttt{\footnotesize zhaoze@seas.upenn.edu}
\And 
Genela Morris \textsuperscript{2,3} \\
\texttt{\footnotesize genelam@tlvmc.gov.il}
\And 
Dori Derdikman \textsuperscript{4} \\
\texttt{\footnotesize derdik@technion.ac.il} \\
\AND
Pratik Chaudhari \textsuperscript{1\textdagger} \\ 
\texttt{\footnotesize pratikac@seas.upenn.edu}
\And 
Vijay Balasubramanian \textsuperscript{5,6\textdagger} \\
\texttt{\footnotesize vijay@physics.upenn.edu}
}

\begin{document}


\maketitle

\vspace{-4ex}
\begin{center}
\footnotesize
\textsuperscript{1}Dept. of Electrical and Systems Engineering, University of Pennsylvania \\
\textsuperscript{2}Tel Aviv Sourasky Medical Center \\
\textsuperscript{3}Gray Faculty of Medical and Health Sciences, Tel Aviv University \\
\textsuperscript{4}Rappaport Faculty of Medicine, Technion – Israel Institute of Technology \\
\textsuperscript{5}Dept. of Physics, University of Pennsylvania
\quad \quad \textsuperscript{6}Santa Fe Institute \\
\textbf{\textsuperscript{\textdagger}Equal contribution}
\end{center}
\vspace{2ex}


\begin{abstract}
Grid cells in the medial entorhinal cortex and place cells in the hippocampus together support spatial navigation. The two regions are reciprocally connected, and there is a chicken-and-egg problem for how both arise and reinforce each other during development. Current computational accounts either derive one type from the other or use network dynamics to model the emergence of one type in isolation. We introduce a unified recurrent network model that instantiates Dale's Law (every neuron is either excitatory or inhibitory), and is trained to predict the next sensory observation from masked previous sensory observations and egocentric motion. To our knowledge, this is the first single-objective model in which grid and place cells co-emerge without supervision of either type, or reliance on pre-existing spatial-cell representations. The two kinds of spatial codes coexist across 1,000 different training configurations, with their balance set by the amount of sensory noise and masking. Without retraining, the network qualitatively reproduces experimentally observed grid fragmentation in hairpin mazes, grid merging after wall removal, lattice alignment across connected rooms, locally ordered 3D fields observed in freely flying bats, as well as the developmental order in which place cells precede grid cells. We interpret these results in terms of two complementary encoding pressures within a single sensory-prediction objective: (1) correcting errors or reconstructing missing components of sensory observations, and (2) prediction of the next sensory state during navigation.  Our results suggest a circuit-level account of the co-emergence of grid and place cells, and experimentally testable predictions for the two kinds of spatial codes. Codes for experiments are available at \footnote{Project Page: \href{https://zhaozewang.github.io/grid-and-place}{https://zhaozewang.github.io/grid-and-place}}\textsuperscript{,}\footnote{Code: \href{https://github.com/grasp-lyrl/grid_and_place}{https://github.com/grasp-lyrl/grid\_and\_place}}.
\end{abstract}


\section{Introduction}

Grid cells in the medial entorhinal cortex (MEC) fire periodically across space, forming a triangular lattice that tiles the environment \cite{haftingMicrostructureSpatialMap2005, moserPlaceCellsGrid2008}. They have been argued to provide an efficient representation of space \cite{wei2015principle}. 
Hippocampal (HC) place cells fire at specific locations \cite{okeefeHippocampusSpatialMap1971, okeefePlaceUnitsHippocampus1976, moserPlaceCellsGrid2015} and remap rapidly between learned environments. While grid cells are thought to emerge from path integration \cite{burakAccuratePathIntegration2009, sorscherUnifiedTheoryComputational2023}, it has been suggested that place cells emerge from encoding sensory experience during navigation \cite{stachenfeldHippocampusPredictiveMap2017, bennaPlaceCellsMay2021, wangTimeMakesSpace2024}. Consistent with these theories, Recurrent Neural Networks (RNNs) trained on path integration develop grid-like activity patterns \cite{baninoVectorbasedNavigationUsing2018, cuevaEmergenceGridlikeRepresentations2018, sorscherUnifiedTheoryComputational2023, chuUnfoldingBlackBox2025, xuConformalIsometryGrid2025}, while RNNs trained to auto-encode sensory observations develop place-like activity patterns \cite{bennaPlaceCellsMay2021, wangTimeMakesSpace2024}. In fact, the RNN in \cite{wangTimeMakesSpace2024} trained, using a similar objective, to predict events across time intervals, also develops units displaying the phenomenology of time cells in the hippocampus \cite{yu2026and}. Together, these studies provide well-established theoretical and computational accounts of each cell type in isolation.

In the brain, grid and place cells coexist within a single bidirectionally connected hippocampal-entorhinal circuit \cite{moserPlaceCellsGrid2008, moserPlaceCellsGrid2015}, and both types provide sparse spatial representations used for self-localization and navigation. In fact, many theories of one cell type rely on the existence of the other.  For example, the authors of \cite{solstadGridCellsPlace2006, rollsEntorhinalCortexGrid2006} proposed that grid cells provide a Fourier-like basis from which place cells arise through Hebbian learning \cite{solstadGridCellsPlace2006, rollsEntorhinalCortexGrid2006}, while \cite{bonnevieGridCellsRequire2013, rennó-costaPlaceGridCells2017} suggested that place cells provide sensory-based error correction that stabilizes grid representations. From the developmental perspective, these potential dependencies create a \emph{\textbf{chicken-and-egg problem}} \cite{bushWhatGridCells2014, morrisChickenEggProblem2023}: if each cell type depends on the other, which arises first? An alternative possibility is that the two types co-emerge.
We will see that simply adding bi-directional interactions to combine RNNs in which grid cells and place cells emerge individually \cite{baninoVectorbasedNavigationUsing2018, cuevaEmergenceGridlikeRepresentations2018, sorscherUnifiedTheoryComputational2023, bennaPlaceCellsMay2021, wangTimeMakesSpace2024} does not lead to co-emergence, perhaps because each type requires distinct recurrent dynamics. 

In this paper, we show how grid and place cells can co-emerge in the hidden layers of a single recurrent network that respects Dale’s Law (each neuron is constrained to be either excitatory or inhibitory), and is trained to predict the next sensory observation from distorted previous sensory observations and relative rotation/movement signals. The network has no prior knowledge of spatial location or pre-existing spatial cell types. This training objective resembles the next-token prediction goal used to train Large Language Models (LLMs), and also reflects the idea that the hippocampal formation implements a predictive map \cite{stachenfeldHippocampusPredictiveMap2017}. Across 1,000 different training configurations representing different amounts of noise, sensory distortion, memory decay rates, and random seeds, we find that grid and place cells robustly co-exist in our network. There are also some regimes where place cells dominate.
The model qualitatively captures experimental observations for both cell types, including responses to wall removal, hairpin-maze fragmentation, connected-room unification, locally ordered grid patterns in freely flying bats in 3D space, and the emergence of place cells before grid cells during development. By linking co-emergence, circuit constraints, and experimental phenomena, our framework offers a mechanistic account of how grid and place codes can arise together to support spatial navigation. Together, these results suggest a principle for spatial representation in the brain, in which predicting the next sensory observation is sufficient to induce complementary place-cell and grid-cell codes from the implicit spatial structure of sensory experience.


\section{From single-cell-type models to co-emergence}

\begin{table}[H]
\caption{Comparison with existing theories of grid and place cell emergence.}
\label{tab:related_work}
\centering
\footnotesize
\setlength{\tabcolsep}{4pt}
\renewcommand{\arraystretch}{1.12}
\begin{tabularx}{\linewidth}{@{}
>{\raggedright\arraybackslash}p{0.3\linewidth}
>{\raggedright\arraybackslash}p{0.16\linewidth}
>{\raggedright\arraybackslash}p{0.16\linewidth}
>{\raggedright\arraybackslash}X
@{}}
\toprule
\textbf{Model class} &
\textbf{Grid cells} &
\textbf{Place cells} &
\textbf{Spatial-cell prior} \\
\midrule
Grid-cell circuit and geometric theories
\cite{burakAccuratePathIntegration2009, burgessOscillatoryInterferenceModel2007, kangGeometricAttractorMechanism2019,
schaefferSelfSupervisedLearningRepresentations2023, xuConformalIsometryGrid2025,chandraEpisodicAssociativeMemory2025, khonaGlobalModulesRobustly2025} &
Mechanistic / \textbf{emergent} &
Not modeled &
Grid-specific circuit or geometric prior \\

\addlinespace[2pt]
Path-integration and place-target RNN models
\cite{baninoVectorbasedNavigationUsing2018, cuevaEmergenceGridlikeRepresentations2018, sorscherUnifiedTheoryComputational2023} &
\textbf{Emergent} &
Simulated target &
Allocentric coordinates or DoG place-cell targets \\

\addlinespace[2pt]
Place-cell emergence models
\cite{bennaPlaceCellsMay2021, wangTimeMakesSpace2024} &
Not modeled &
\textbf{Emergent} &
No grid-cell prior \\

\addlinespace[2pt]
Successor-representation and predictive-map theories
\cite{stachenfeldHippocampusPredictiveMap2017} &
Derived / spectral &
Derived / predictive &
Explicit state graph or transition structure \\

\addlinespace[2pt]
Grid-place interaction and dependency models
\cite{solstadGridCellsPlace2006, rollsEntorhinalCortexGrid2006, bonnevieGridCellsRequire2013, rennó-costaPlaceGridCells2017} &
Predefined &
Predefined / mechanistic &
Pre-existing grid or place cells \\

\addlinespace[2pt]
System-level grid-place models
\cite{wangREMIReconstructingEpisodic2025, chandraEpisodicAssociativeMemory2025} &
Predefined &
Predefined / constructed &
Pre-existing spatial-cell representations \\

\bottomrule
\end{tabularx}
\end{table}

Existing models of the entorhinal-hippocampal spatial code fall into  distinct families (Table~\ref{tab:related_work}):
grid-cell circuit and RNN models that rely on spatial supervision or grid-specific priors without modeling place-cell emergence
\cite{burakAccuratePathIntegration2009,
burgessOscillatoryInterferenceModel2007, cuevaEmergenceGridlikeRepresentations2018,
kangGeometricAttractorMechanism2019,
sorscherUnifiedTheoryComputational2023, schaefferSelfSupervisedLearningRepresentations2023,xuConformalIsometryGrid2025, khonaGlobalModulesRobustly2025},
place-cell emergence models that obtain localized fields from sensory experiences but do not produce grids \cite{bennaPlaceCellsMay2021, wangTimeMakesSpace2024},
and interaction or system-level models that assume that one or both cell types already exist \cite{solstadGridCellsPlace2006, rennó-costaPlaceGridCells2017, chandraEpisodicAssociativeMemory2025}.  Our goal in this work is to study how both representations can co-emerge in a single connected circuit.

We will work with a shared recurrent network derived from continuous-time neural dynamics that has also been used in previous work \cite{cuevaEmergenceGridlikeRepresentations2018, baninoVectorbasedNavigationUsing2018, sorscherUnifiedTheoryComputational2023, schaefferSelfSupervisedLearningRepresentations2023, wangTimeMakesSpace2024, xuConformalIsometryGrid2025}. The discrete-time update is given by
\begin{equation}
h_{i,t+1} = (1 - \alpha)\, h_{i,t}
+ \alpha \Big(
\textstyle\sum_{j=1}^{N} W_{ij}\,\phi(h_{j,t})
+
\sum_{k=1}^{d_{\text{in}}} B_{ik}\, u_{k,t}
+
b_i
\Big),
\label{eq:rnn_discrete}
\end{equation}
where \(h_{i,t}\) is the membrane potential of neuron \(i\) at time  \(t\), the recurrent weight from neuron \(j\) to  \(i\) is denoted by \(W_{ij}\), the input projection by \(B\), the external input to the network is \(u_t \in \mathbb{R}^{d_{\mathrm{in}}}\), the bias is \(b_i\), and the firing-rate nonlinearity (ReLU or softplus) is \(\phi(\cdot)\).

\subsection{Direct composition of  single-cell-type models}

\begin{figure}[!htbp]
\centering
\includegraphics[width=\textwidth]{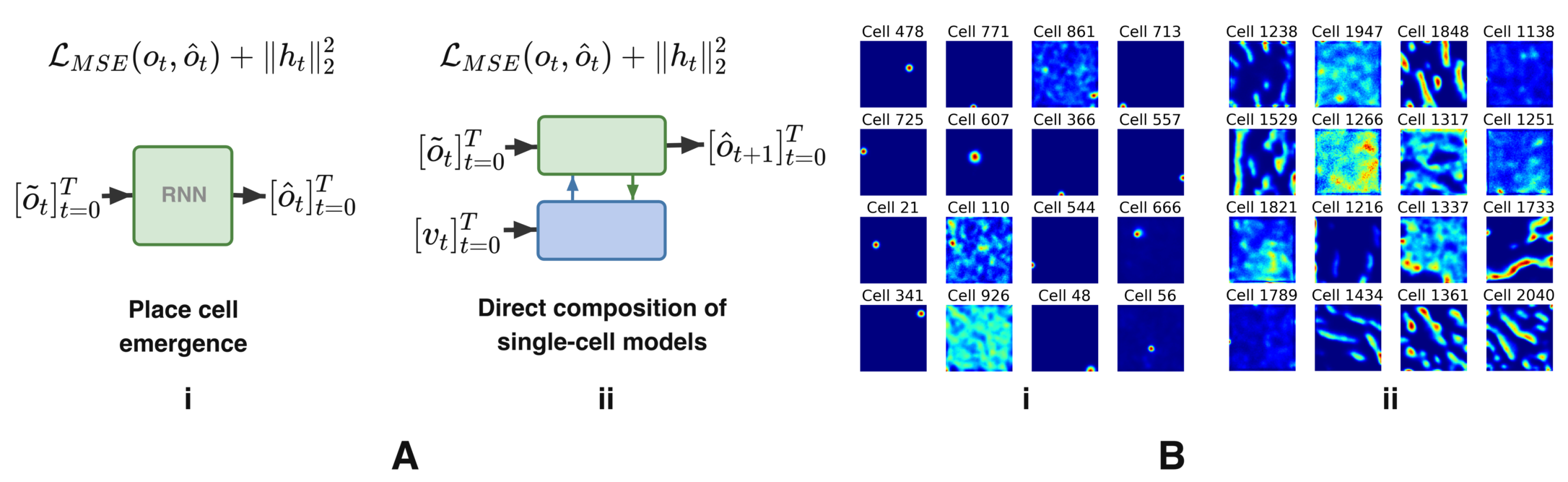}
\caption{\textbf{A.}~(i)~An RNN that reconstructs  \(\hat{o}_t\) from corrupted observations \(\tilde{o}_t\) with an $\ell_2$ penalty on firing rates leads to place-like cells \cite{wangTimeMakesSpace2024}. (ii)~Composition of place and grid cell emergence models within a single network, via bidirectional connections, extends the place cell model in \cite{wangTimeMakesSpace2024} to have an additional input layer, which receives velocity input like the grid cell models in  \cite{cuevaEmergenceGridlikeRepresentations2018, sorscherUnifiedTheoryComputational2023}.
\textbf{B.}~Emergent cells in the composed model. (i)~Place-like cells in the sensory-driven region. (ii)~Irregular stripe- or multi-peaked fields in the ``free'' region without sensory input. The triangular lattices of grid cell responses are absent.
}
\label{fig:previous_frameworks}
\end{figure}

We first check whether models in which each cell type emerges individually can be combined to get co-emergence of grid and place cells.
Sensory denoising or reconstruction during navigation in RNNs (Fig.~\ref{fig:previous_frameworks}Ai)
can produce place-cell-like representations \cite{wangTimeMakesSpace2024}. Similarly,
motion-conditioned prediction of spatial location by RNNs can produce grid-cell-like representations \cite{cuevaEmergenceGridlikeRepresentations2018, baninoVectorbasedNavigationUsing2018, sorscherUnifiedTheoryComputational2023}.
In \cite{cuevaEmergenceGridlikeRepresentations2018}, the network was trained to predict allocentric coordinates, while in \cite{baninoVectorbasedNavigationUsing2018,sorscherUnifiedTheoryComputational2023}, location was derived from supervised place cell activity, modeled in \cite{sorscherUnifiedTheoryComputational2023} as hand-coded Difference of Gaussian (DoG) spatial filters.
Since grid-cell models learn by predicting place-cell activity, it is natural to compose the two models by replacing hand-designed place targets with emergent place-like representations.

We start with the emergent place-cell network in \cite{wangTimeMakesSpace2024} (green in Fig.~\ref{fig:previous_frameworks}Aii), and recurrently connect another network that receives motion input (blue in Fig.~\ref{fig:previous_frameworks}Aii).
Motivated by the successor representation framework for place cells \cite{stachenfeldHippocampusPredictiveMap2017}, we modify the model of \cite{wangTimeMakesSpace2024} so that it predicts the next sensory input from a noisy and masked current sensory observation.
In Fig.~\ref{fig:previous_frameworks}Aii, we should expect this information to propagate to the blue region via back-propagation and induce units there to predict the place cell activity. This is similar to the direct supervision in \cite{cuevaEmergenceGridlikeRepresentations2018, baninoVectorbasedNavigationUsing2018, sorscherUnifiedTheoryComputational2023}. Our trained network successfully reconstructs noisy or masked sensory experiences along simulated trajectories (details in Supp.~\ref{sup:trajectory}). It develops place-cell-like representations in the green region (Fig.~\ref{fig:previous_frameworks}Bi). But the blue region produces neurons with irregular activity stripes or firing fields with multiple peaks (Fig.~\ref{fig:previous_frameworks}Bii). It does not exhibit triangular lattices characteristic of grid cells.

\subsection{Key ingredients for co-emergence}
\label{sec:composition_challenges}
Direct composition of networks with emergent place and grid fields is difficult not only because parameter tuning may be delicate, but also because mechanisms that produce each cell type separately impose distinct constraints on the recurrent circuit as described below. Additionally, in the brain there are constraints imposed by the biophysics of neurons and by the way in which animals experience the world relative to self. 

\textbf{Temporal update constraint.}
Different spatial codes may require different temporal update dynamics. In our recurrent network, the timescale is controlled by the decay rate \(\alpha\) in Eq.~\ref{eq:rnn_discrete}. Smaller \(\alpha\) produces slower, more persistent dynamics and larger \(\alpha\) produces faster updates. Place-cell-like denoising benefits from slower dynamics, because partial or corrupted sensory cues must be integrated toward a consistent network state. Grid-cell-like transitions instead require that the network preserves the current state when the animal is stationary, but rapidly updates the state when the animal moves. This cannot be achieved by assigning different fixed decay rates \(\alpha\) to different neurons or regions: a fixed rate sets a constant decay timescale independent of behavioral state. What is needed is effectively a gated update, in which the recurrent state is stable in the absence of motion but can be rapidly reconfigured by motion-dependent input. Thus, the difficulty is not only that place and grid codes prefer different timescales, but that grid-like codes require state-dependent switching between persistence and fast update.

\textbf{Structural constraint.} A simple combination like the one in Fig.~\ref{fig:previous_frameworks}Aii imposes a strong structural constraint: one population is implicitly assigned to the hippocampal role by receiving sensory inputs to support place cells, while the other is implicitly assigned to the MEC role by receiving motion input, and is designated as the putative grid-cell population. This pre-specified division may impede the dynamics required for co-emergence, since each population is strongly shaped by its designated input stream rather than self-organizing under a shared predictive objective. Furthermore, the conventional separation into hippocampus and MEC could simply be a description of functional and structural distinctions that have emerged in a single network that performs a shared task. Thus, pre-assigning separate hippocampal and MEC-like populations may obscure the required self-organization and functional specialization through which grid-like and place-like codes emerge.

\textbf{Self-motion input constraint.} Existing grid-cell RNN models take motion input in terms of allocentric displacement or velocity \cite{baninoVectorbasedNavigationUsing2018, cuevaEmergenceGridlikeRepresentations2018, sorscherUnifiedTheoryComputational2023}. This setup assumes that movement is already expressed in an external spatial reference frame. However, during navigation, self-motion cues are more readily available in body-centered forms, such as relative rotation and speed from vestibular, proprioceptive, and motor-related signals, and are presumably converted into allocentric movement signals by network mechanisms that may also exploit external cues and path integration.

\textbf{Biophysical constraint.} In typical RNNs, including the model above, the connection weight matrix is unconstrained.  But in the brain neurons satisfy Dale's Law: the outgoing projections of a given neuron are all excitatory (positive weights in an RNN) or inhibitory (negative weights in an RNN)

\begin{figure}[!htpb]
\centering
\includegraphics[width=0.92\textwidth]{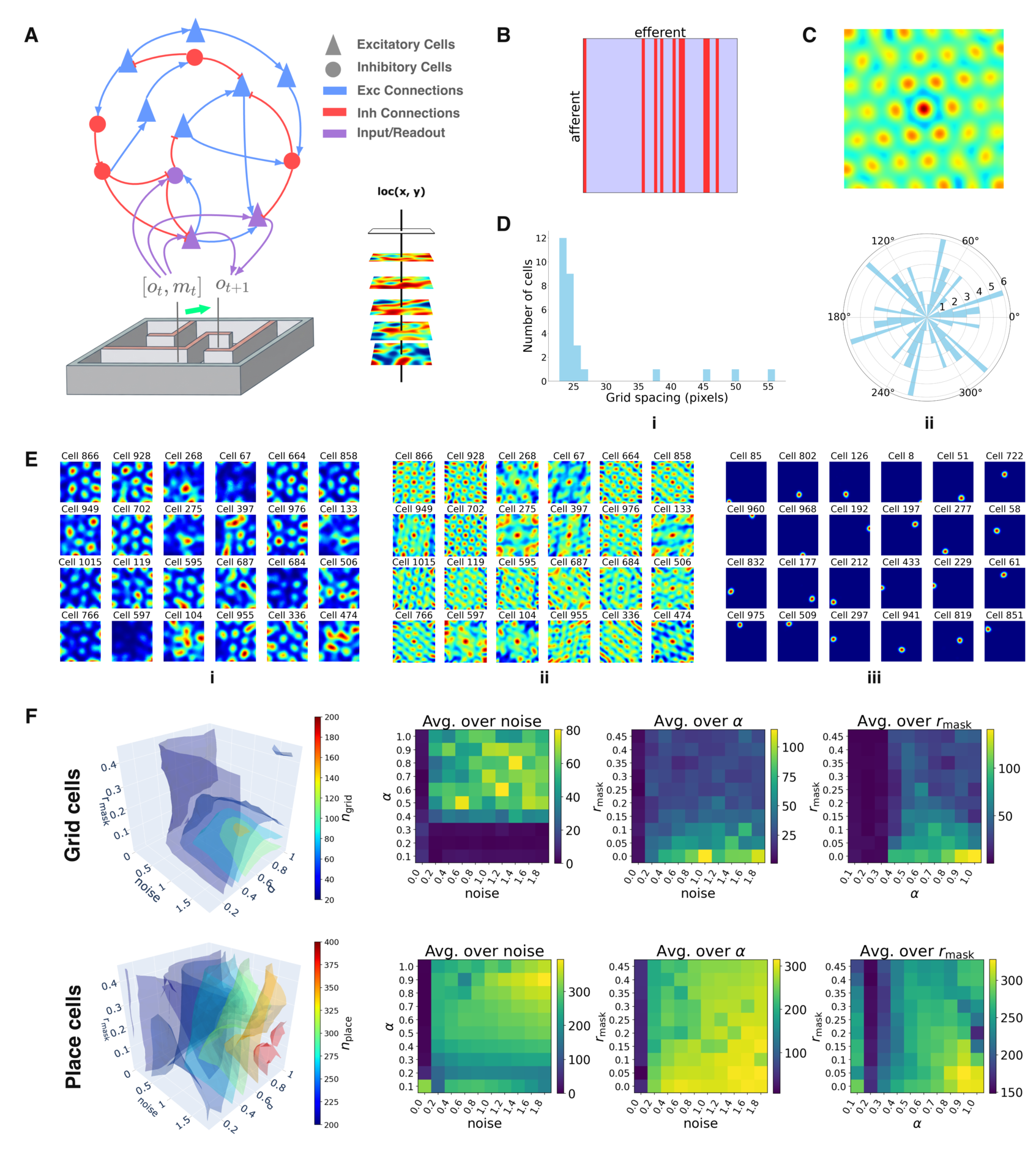}
\caption{
\textbf{A.}~Architecture of the RNN. Excitatory (triangles) and inhibitory (circles) neurons interact through recurrent connections constrained by Dale's law. The hidden layer is split into input-driven neurons, which receive noisy masked sensory inputs $\tilde{o}_t$ and egocentric motion input $m_t$, and recurrently driven neurons, which receive no direct external input. The motion signal consists of relative rotation from the previous timestep, represented as a unit vector, and a scalar speed normalized to \([0,1]\). 
\textbf{B.}~Illustration of the recurrent weight matrix. Rows index efferent neurons and columns index afferent neurons; 80\% of neurons are excitatory (blue, non-negative outgoing weights) and 20\% inhibitory (red, non-positive outgoing weights).
\textbf{C.}~Population-averaged spatial autocorrelogram of cells classified as grid cells. Each autocorrelogram was peak-normalized before averaging, so the panel reflects the shared lattice structure instead of being dominated by high-rate cells. The six satellite peaks around the center summarize the population-level triangular periodicity.
\textbf{D.}~Grid spacing and orientation statistics across emergent grid cells. (i)~Distribution of grid spacings, measured in pixels from the spatial autocorrelogram peaks. (ii)~Distribution of grid orientations. Orientation is measured as the angle between the horizontal axis of the rate map and the nearest grid-lattice axis in the spatial autocorrelogram, modulo \(60^\circ\) to account for hexagonal symmetry. The polar plot repeats this distribution to extend across the circle visualization purposes.
\textbf{E.}~Example single-cell ratemaps. (i)~Grid cell response field from the free region. (ii)~Spatial autocorrelograms of these grid cells, showing triangular lattice symmetry. (iii)~Place cell response fields from the sensory-driven region.
\textbf{F.}~Parameter sweep over noise level \(\sigma_n\), decay rate \(\alpha\), and masking fraction \(r_{\text{mask}}\). Top row: grid cell counts shown as isosurfaces (left) and marginal heatmaps averaged across different parameters (right). To aid visualization, cell counts were smoothed across adjacent parameter settings before extracting the isosurfaces. Bottom row: place cell counts in the same layout.
}
\label{fig:main_results}
\end{figure}


\section{Co-emergence of grid and place cells}
\label{sec:main_results}

These constraints described above have corresponding circuit-level implementations. For the \textbf{\emph{temporal update constraint}}, recurrent inhibition can support rapid state reconfiguration \cite{tsodyksParadoxicalEffectsExternal1997, vanVreeswijk1996}, while the decay term allows activity to decay gradually over time.  To address the \textbf{\emph{structural constraint}}, we do not pre-assign separate hippocampal and MEC populations. Instead, we treat the circuit as a single recurrent population and divide neurons in terms of their external input. \emph{Input-driven neurons} receive direct sensory and motion input. \emph{Recurrently driven neurons} receive no direct external input and are shaped only through recurrent interactions with the rest of the network. This design avoids imposing a predefined grid/place or HC/MEC connectivity structure, allowing potential grid- and place-like subpopulations to self-organize under a shared predictive objective. Finally, to address the \textbf{\emph{self-motion input constraint}}, we provide motion as egocentric self-motion signals----rotation relative to previous heading and speed---rather than allocentric displacement or velocity. This forces the network to learn spatial codes from egocentric sensory experience and body-centered motion cues without assuming movement expressed in a global spatial reference frame.  To implement the \textbf{\emph{biophysical constraint}}, we require the recurrent weight matrix to satisfy Dale's Law by assigning a fixed sign to the weights of outgoing synapses of each neuron throughout training. This is a biologically realistic recurrent structure through which inhibitory feedback can shape dynamics. We also fix the bias term in Eq.~\ref{eq:rnn_discrete} to zero, for two reasons. A learnable bias lacks a clear biological counterpart, and it can suppress firing rates via a global offset and thus bypass inhibitory feedback.

We implement these conditions in a recurrent network with $N=2048$ neurons. Half of the neurons receive direct external input, while the remaining half are only driven through recurrent connections. At initialization, 80\% of neurons are deemed excitatory (non-negative outgoing weights), and 20\% are deemed inhibitory (non-positive outgoing weights), consistently with typical proportions in hippocampus and cortex, and this constraint is enforced during training (see Supp.~\ref{sup:dale}). The precise ratio between input-driven and recurrently driven neurons does not qualitatively affect the results presented below (see Supp.~\ref{sup:region_ratio}).

At each step the network receives a masked sensory observation $\tilde{o}_t$ (a fraction $r_{\text{mask}}$ dimensions are masked) and an egocentric motion signal $m_t$ consisting of relative rotation and a scalar speed normalized to \([0,1]\), and predicts the next observation $\hat{o}_{t+1}$. The network is trained to minimize the mean square error between the predictions and ground-truth unmasked observations. Sensory observations are generated over a 220\,cm $\times$ 220\,cm arena so that nearby locations have similar observations (details in Supp.~\ref{sup:sensory}). After each recurrent update, we inject additive Gaussian noise with amplitude controlled by a parameter $\sigma_n$ and modulated by the firing rate (Supp. Eq.~\ref{eq:fr_noise}), so that neurons that are more active experience larger perturbations. All biases are set to zero.

\subsection{Robust co-emergence of grid and place cells}
\label{sec:coemergence_robustness}

After training, the network develops grid- and place-like representations in the recurrently driven and input-driven neurons, respectively (Fig.~\ref{fig:main_results}E.i,iii). At the final analysis step, our classification criteria counted 58 grid-like units, predominantly among recurrently driven neurons, using a grid-score threshold of \(>0.3\) (see Supp.~\ref{sup:cell_classification}), and 307 place-like units across both input-driven and recurrently driven neurons (see Supp.~\ref{sup:cell_classification} for place-cell classification methods). The emerged grid cells exhibit multiple firing fields arranged in a triangular lattice with mean spacing 25.6\,cm (Fig.~\ref{fig:main_results}D.i). Grid orientation is measured from the spatial autocorrelogram as the angle between the horizontal axis of the rate map and the nearest grid-lattice axis modulo \(60^\circ\) to account for triangular symmetry (Fig.~\ref{fig:main_results}D.ii). The identified grid cells show the expected triangular symmetry in their autocorrelograms (Fig.~\ref{fig:main_results}E.ii). Across \(10\) random-seed replicates at the same training step, these criteria yielded \(30.0 \pm 19.2\) grid-like units and \(299.4 \pm 4.9\) place-like units (mean \(\pm\) s.d.). The greater seed-to-seed variability in the grid-like population may partly reflect the sensitivity of existing grid-cell classification metrics. Consistent with this interpretation, when the grid-score threshold was relaxed to \(>0.1\), 542 cells showed grid-like periodic firing fields, including cells with locally triangular fields that were distorted near arena boundaries and therefore did not form a regular global \(60^\circ\) pattern. Nevertheless, all networks trained with different random seeds robustly developed units with periodic firing fields that qualitatively resembled grid cells.

We vary three parameters to test the robustness of the model: the activity decay rate $\alpha \in [0.1, 1.0]$, noise level $\sigma_n \in [0, 0.45]$ in the recurrent network, and input masking fraction $r_{\text{mask}} \in [0, 0.9]$, over a $10 \times 10 \times 10$ grid (details in Supp.~\ref{sup:parameter_sweep}). We train each model for 20,000 steps with a \emph{different random seed} and identify cell types using the criteria in Supp.~\ref{sup:cell_classification}. For this sweep, we use the more permissive grid-score threshold of 0.1, which better captures how grid-like periodic structure varies across training conditions. Fig.~\ref{fig:main_results}F shows that place cells emerge robustly across nearly all settings. Grid-like periodicity preferentially emerges at higher \(\sigma_n\), and is modulated more weakly by the decay rate \(\alpha\). Higher masking ratio \(r_{\text{mask}}\) causes place cells to also appear among recurrently driven neurons, while grid-like periodicity fails to emerge.

To better understand which architectural components support the co-emergence of grid and place cells, we modify individual components while keeping the other settings unchanged. First, we remove the \textbf{Dale's Law constraint}, allowing all recurrent weights to be freely learned. This reduces the number of grid cells among recurrently driven neurons from 58 to 3 under the grid-score threshold of 0.3. The remaining cells develop multi-peaked or stripe-like irregular fields without triangular lattice structure, similar to the outcome of the direct composition model in Sec.~\ref{sec:composition_challenges}. We next test the role of \textbf{inhibitory feedback} by making the bias term \(b_i\) in Eq.~\ref{eq:rnn_discrete} learnable. Our co-emergence model fixes \(b_i=0\) because a learnable bias lacks a clear biological counterpart and can suppress firing rates through a global offset, thereby bypassing inhibitory feedback. Under this modification, no cells pass our standard grid-cell criterion of mean firing rate above 0.1\,Hz and grid score above 0.3, whereas place cells still emerge among the input-driven neurons. Finally, we vary the \textbf{motion input} by training networks with allocentric motion input, and also with motion input removed completely. Grid- and place-like cells still emerge in both settings, suggesting that explicit motion input is not required when consecutive sensory observations already contain smooth transition information along trajectories.


\begin{figure}[!htpb]
\centering
\includegraphics[width=\textwidth]{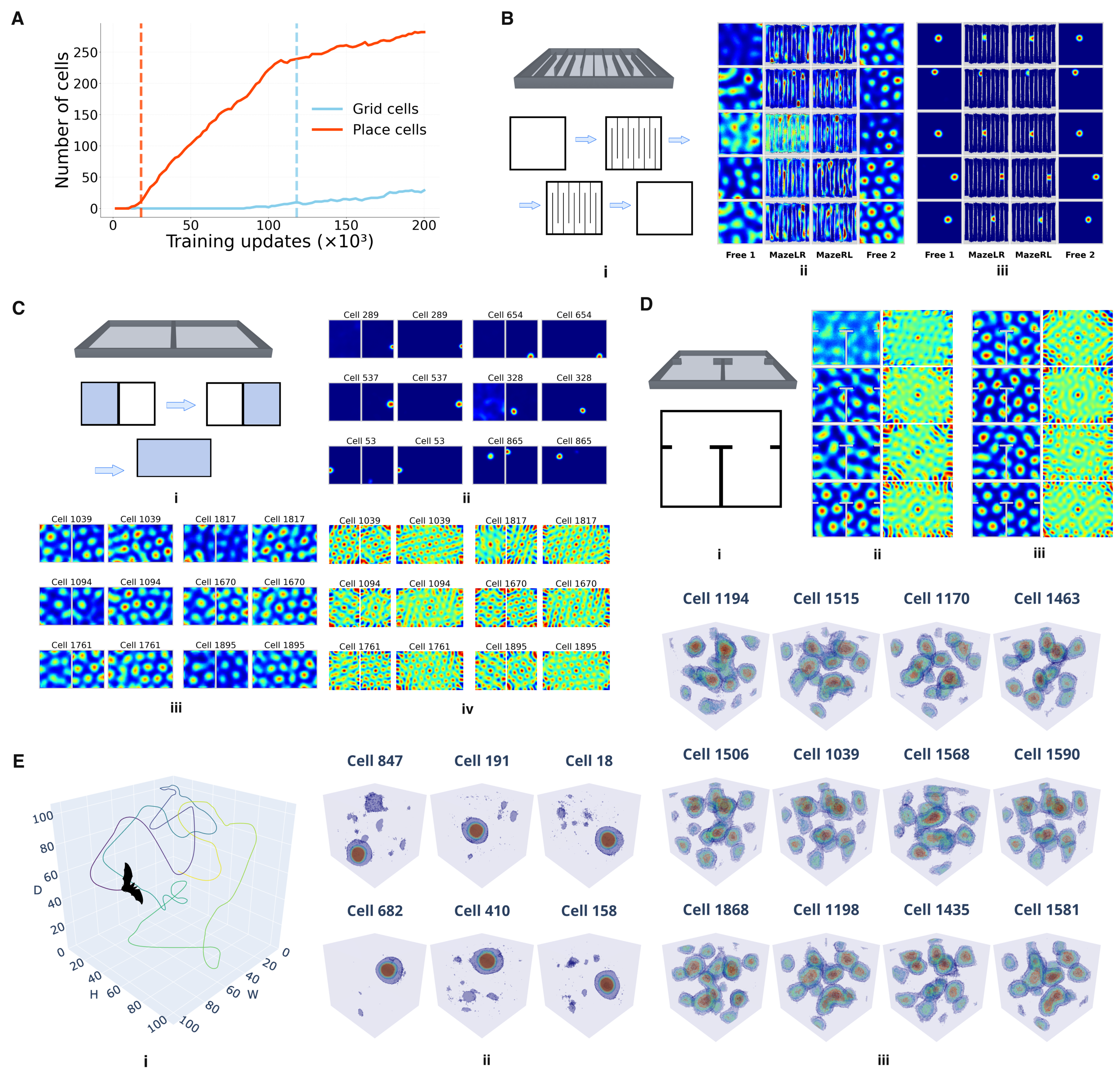}
\caption{\textbf{A.}~Developmental order of emerged representations. Grid-cell counts (cyan) and place-cell counts (orange) are shown across training. Place cells emerge earlier than grid cells; dashed lines mark the first step at which each cell type exceeds a count of~5.
\textbf{B.}~Hairpin maze \cite{derdikmanFragmentationGridCell2009}. (i)~Schematic of the open-field and hairpin trials. The animal first explores an open arena, then traverses the hairpin maze under constrained trajectories, and finally returns to the open arena. (ii)~Grid-cell ratemaps fragment across corridors when movement is constrained, but recover coherent hexagonal patterns in the open arena. (iii)~Place cells remain spatially stable across conditions.
\textbf{C.}~Two-room wall removal. (i)~Schematic of two rooms separated by a wall and then merged after wall removal. The animal sequentially explores the left room, the right room, and the merged arena. (ii--iv)~Ratemaps and autocorrelograms across trials. For compactness, ratemaps from the left- and right-room trials are plotted together with a wall in between, while even columns show responses after wall removal for the same cell. (ii)~Place-cell firing centers remain stable before and after wall removal. (iii)~Grid cells develop independent hexagonal patterns in each room when the wall is present, and merge into a unified lattice after wall removal. (iv)~Autocorrelograms show the same transition from two independent lattices to one unified lattice.
\textbf{D.}~Connected rooms \cite{carpenterGridCellsForm2015}. (i)~Schematic of two rooms connected by a corridor. The animal can freely traverse between rooms through the corridor. (ii--iii)~Grid-cell ratemaps and autocorrelograms early and late in training. Early in training, grid fields are locally ordered within each room but remain fragmented across the connected environment. Later in training, some cells develop more globally ordered patterns, with grid fields becoming aligned across both rooms and the connecting corridor.
\textbf{E.}~Three-dimensional volumetric traversal. (i)~Example 3D trajectory. (ii)~Place-like cells develop localized 3D firing fields. (iii)~Grid-like cells develop locally ordered volumetric fields, consistent with partially ordered grid responses recorded in flying bats.
\label{fig:multi_compartment}
}
\end{figure}

\section{Properties of the emergent grid and place cells}
\label{sec:cross_environment_experiments}

Next we test whether the co-emergent grid/place cells exhibit properties of their biological counterparts.

\textbf{Developmental order (Fig.~\ref{fig:multi_compartment}A).}  
Across parameters,  place cells emerge before grid cells during training (Fig.~\ref{fig:multi_compartment}A), consistently with recordings during rodent development \cite{willsDevelopmentHippocampalCognitive2010, langstonDevelopmentSpatialRepresentation2010, muessigDevelopmentalSwitchPlace2015}. 

\textbf{Hairpin maze (Fig.~\ref{fig:multi_compartment}B).} We replicate the experiment of \cite{derdikmanFragmentationGridCell2009}, in which grid cells were recorded as rodents first explored an open field,  then traversed an imposed hairpin maze, and then returned to the open field. We model this setting by using the same sensory noise field across environments, slightly modified to respect the maze walls (details in Supp.~\ref{sup:hairpin}), so that corresponding spatial locations in the maze and open field provide similar sensory observations while local sensory similarity in the hairpin maze follows the corridor structure. In the open arena, grid cells exhibit coherent triangular patterns. In the hairpin maze, where movement is constrained to narrow corridors, grid patterns fragment across corridors and the spatial autocorrelograms become banded rather than triangular. Upon returning to the open field, triangular firing patterns gradually recover as training continues in the open field. 
These results recapitulate \cite{derdikmanFragmentationGridCell2009}.
Place-cell firing centers, which were not reported in \cite{derdikmanFragmentationGridCell2009},  remain spatially aligned between the hairpin maze and the open arena, such that cells active at a given location in the open arena are also active at the corresponding location in the maze. Hence, our model makes a prediction: in the conditions of \cite{derdikmanFragmentationGridCell2009}.  place-cell firing centers should remain spatially aligned between the hairpin maze and the open arena, even as grid-cell maps fragment.

\textbf{Two-room wall removal (Fig.~\ref{fig:multi_compartment}C).}  We train the network for 60,000 steps in room~1 and 60{,}000 steps in room~2 while the rooms are separated by a wall, followed by 80{,}000 steps in the combined arena after wall removal. This tests whether the learned spatial code treats the two rooms as separate environments or as one continuous space. When the wall is present, grid cells develop independent triangular lattice patterns in the two compartments, visible in both ratemaps and autocorrelograms. After wall removal, grid patterns merge into a unified lattice spanning both rooms, while place cells maintain localized and spatially stable firing fields. This result is as a prediction of our model: after wall removal, grid-cell maps should reorganize into a unified lattice spanning the combined arena, while place-cell firing centers should remain spatially localized and stable.

\textbf{Connected rooms (Fig.~\ref{fig:multi_compartment}D).} Following \cite{carpenterGridCellsForm2015}, we connect two rooms by a corridor that the animal can freely traverse. The original experiment tested whether grid cells form local maps within each room or a global map across connected spaces. This setting is related to the wall-removal experiment, but is more challenging because the rooms are linked only indirectly through a corridor, requiring spatial periodicity to align across a longer traversal path. We therefore compare grid-cell responses early and late in training. Early in training, grid cells show fragmented, locally triangular lattice patterns within each room (Fig.~\ref{fig:multi_compartment}D.ii). Later in training, some cells develop global patterns, with grid fields and periodicity becoming ordered across the connected environment despite the intervening walls, resembling findings in \cite{carpenterGridCellsForm2015}. Although not all cells form globally ordered lattices, grid cells in both groups preserve locally ordered structure, with local grid orientation gradually drifting across locations. This suggests that the network tends toward global alignment, but errors accumulated over long indirect traversal paths can prevent perfectly coherent global lattices.

\textbf{Three-dimensional traversal (Fig.~\ref{fig:multi_compartment}E).} Finally, we train the network in a 3D environment in which the agent moves freely through the full volume. The trajectory is generated by the same smoothed random-walk procedure with boundary avoidance used in 2D environments (Supp.~\ref{sup:trajectory}). We find that place cells develop clear localized 3D firing fields. Grid cells also develop structured volumetric firing fields, but these fields are locally ordered rather than organized as a perfect global lattice, resembling locally ordered grid-like responses observed in freely flying bats \cite{ginosarLocallyOrderedRepresentation2021}. 
The local ordering that we see may reflect the difficulty of uniformly traversing a 3D volume. As discussed in the next section, grid formation in our network its trajectory-dependent; the difficulty of uniformly traversing and sampling a 3D volume may be leading to the irregularity of the emergent grid fields.



\section{Discussion}

Our goal in this paper is to build a conceptual framework for understanding the functional logic of the circuits in the animal brain that facilitate spatial navigation. The crux of our model is the predictive objective---in order to navigate, animals have to be able to predict the expected sensory input that will result from their current motion. This predictive capability underpins goal-directed planning.  We instantiate this objective in an RNN model of the hippocampal formation with a few key biological constraints. Chief among these is Dale's Law---a neuron has either excitatory or inhibitory projections, but not both. We also include firing rate modulated noise, and relative proportions of excitation and inhibition similar to those in the brain.  We do not seek here to develop a detailed description of the biological circuits, but rather to reveal the essential constraints and architectural motifs that are required.
As such, rather than separating the network into a putative MEC (where grid cells would be found) receiving motion input and a putative hippocampus (where place cells would be found) receiving sensory input, we consider an architecture in which one component receives external input (sensory and motion) and another only has recurrent connections. This allows us to better understand the dynamics that lead to the formation of the two cell types. After learning, the network robustly develops place-cell-like units in the component receiving sensory input and grid-cell-like units in the other component. The emergent place and grid cells reproduce a number of experimentally observed phenomena, including earlier development of place fields, fragmentation of grid fields in hairpin mazes,  formation of global grid representations in connected environments, and 3D grid fields with local but not global order.  We model predicts new phenomena, e.g, place fields remain stable if a hairpin maze is introduced into an open field environment, or if a wall between two rooms is removed. The simplicity of our model makes it possible to easily predict the consequences in new experimental regimes.

Why do place and grid representations appear in the trained network? The animal's sensory experience traces a continuous trajectory on a manifold within high-dimensional sensory space, which is constrained to have the same low dimension as the physical space being explored. In this context,  predicting the next sensory observation can be decomposed into two distinct problems: (i) \textit{\textbf{off-manifold contraction}}, and (ii) \textit{\textbf{on-manifold transition}}.  In detail, masking observations perturbs them away from the manifold of valid sensory states. The network pulls the perturbed state back to the manifold. The mechanism for doing so has to be a spatially localized attractor basin -- otherwise it might correct to the wrong location. Phenomenologically, the authors of \cite{wangREMIReconstructingEpisodic2025} showed that this process indeed occurs in networks like ours and the networks show place-cell-like responses. This suggests that place cell representations can be thought of as coordinates specifying the manifold of valid sensory experiences within the space of all possible sensory signals. Sensory observations change smoothly as the animal moves along a trajectory. To predict the next observation, its circuit should use the current motion to determine the next location on the sensory manifold. If the spatial location were represented in some local chart, it could be updated using the current velocity by a recurrent circuit.  Grid cells can be interpreted as maintaining such a local chart \cite{fiete2008grid, bushUsingGridCells2015, wei2015principle}.  This update can be fed back to the circuit maintaining the sensory manifold (which contains the place cells). Altogether, this predictively updates the location on the sensory manifold.

Our model aligns with several existing theories of grid and place cells. \emph{\textbf{Continuous-attractor models}} show how recurrent circuit dynamics maintain and update an internal spatial state during movement \cite{burakAccuratePathIntegration2009}. Our model similarly treats grid cells as supporting motion-driven transitions, but explains how they can arise in a Dale's Law-constrained recurrent circuit trained by next-observation prediction. Our model also follows existing \emph{\textbf{path-integration RNN models}} \cite{baninoVectorbasedNavigationUsing2018,cuevaEmergenceGridlikeRepresentations2018,sorscherUnifiedTheoryComputational2023}, but addresses two assumptions in these models. First, instead of assuming Difference-of-Gaussian (DoG) place-cell targets  \cite{sorscherUnifiedTheoryComputational2023}, our model treats place cells as emergent from an excitatory-inhibitory network.  Second, instead of requiring allocentric motion inputs, grid cells emerge in the network with egocentric inputs consistent with the experience of animals, or even without explicit motion signals.  In the latter case,  the network can infer motion information directly from changes in sensory signals. On the other hand, \emph{\textbf{self-supervised grid-cell models}} \cite{schaefferSelfSupervisedLearningRepresentations2023, xuConformalIsometryGrid2025} train grid representations using explicit geometric constraints, such as contrastive similarity between nearby positions \cite{schaefferSelfSupervisedLearningRepresentations2023} or local distance and angle preservation \cite{xuConformalIsometryGrid2025}. Our model suggests a possible source for these geometry-preserving constraints: adjacent locations produce similar sensory experience, so next-observation prediction implicitly encourages nearby states to have related representations. Finally, our theory builds directly on \emph{\textbf{episodic-memory theories}} of place cells \cite{bennaPlaceCellsMay2021, wangTimeMakesSpace2024}, while connecting them to \emph{\textbf{successor-representation}} accounts \cite{stachenfeldHippocampusPredictiveMap2017}. In our model, predictive structure is supported not only by place-like state specificity but also by grid-like action-conditioned transitions along the sensory manifold.

\subsection{Limitations and Future Directions}

Grid cells in the MEC are known to be organized in a hierarchy of discrete modules \cite{stensola2012entorhinal}.  Our network only produced a single module in which the grid cells have similar periods. It is possible to encode space with such a distribution of periods \cite{fiete2008grid}, but it is known that the discrete scaling hierarchy of the type observed in \cite{stensola2012entorhinal} is more efficient \cite{wei2015principle}.  Perhaps such modules would be obtained if our network were to have multiple layers or some further regularization that incentivizes representational efficiency \cite{kangGeometricAttractorMechanism2019,khonaGlobalModulesRobustly2025}.  Furthermore, experiments shown that if an enclosure is suddenly stretched or shrunk, grid fields appear to distort correspondingly \cite{barry2007experience,stensolaShearinginducedAsymmetryEntorhinal2015, krupicGridCellSymmetry2015}. There is evidence that this apparent distortion arises from time-averaging phase-shifts in the grid pattern that occur when the animal approaches a boundary in the distorted direction, possibly because of interaction between grid cells and border cells \cite{hardcastle2015environmental, keinathEnvironmentalDeformations2018}.  It would be interesting to test whether our network reproduces such distortion effects, and/or whether it contains emergent border cells in addition to the place and grid cells that we have analyzed.  In fact, a preliminary analysis of our network revealed units that showed border cell-like responses, and others that show responses resembling allocentric head direction cells, which are known to be present in the MEC and hippocampus.  A goal for the future should be to understand the conditions and constraints under with the repertoire of cell types seen in these brain structures is reproduced.

\begin{ack}
The study was supported by NIH CRCNS grant 1R01MH125544-01 and in part by the NSF and DoD OUSD (R\&E) under Agreement PHY-2229929 (The NSF AI Institute for Artificial and Natural Intelligence). Additional support was provided by the United States–Israel Binational Science Foundation (BSF). PC was supported in part by grants from the National Science Foundation (IIS-2145164, CCF-2212519).
\end{ack}

\clearpage
\bibliographystyle{unsrtnat}
\bibliography{grid,others}

@article{barry2007experience,
  title={Experience-dependent rescaling of entorhinal grids},
  author={Barry, Caswell and Hayman, Robin and Burgess, Neil and Jeffery, Kathryn J},
  journal={Nature neuroscience},
  volume={10},
  number={6},
  pages={682--684},
  year={2007},
  publisher={Nature Publishing Group US New York}
}

@article{stensola2012entorhinal,
  title={The entorhinal grid map is discretized},
  author={Stensola, Hanne and Stensola, Tor and Solstad, Trygve and Fr{\o}land, Kristian and Moser, May-Britt and Moser, Edvard I},
  journal={Nature},
  volume={492},
  number={7427},
  pages={72--78},
  year={2012},
  publisher={Nature Publishing Group UK London}
}

@article{fiete2008grid,
  title={What grid cells convey about rat location},
  author={Fiete, Ila R and Burak, Yoram and Brookings, Ted},
  journal={Journal of Neuroscience},
  volume={28},
  number={27},
  pages={6858--6871},
  year={2008},
  publisher={Society for Neuroscience}
}

@article{wei2015principle,
  title={A principle of economy predicts the functional architecture of grid cells},
  author={Wei, Xue-Xin and Prentice, Jason and Balasubramanian, Vijay},
  journal={Elife},
  volume={4},
  pages={e08362},
  year={2015},
  publisher={eLife Sciences Publications, Ltd}
}

@article{yu2026and,
  title={When and Where: A Model Hippocampal Network Unifies Formation of Time Cells and Place Cells},
  author={Yu, Qiaorong S and Wang, Zhaoze and Balasubramanian, Vijay},
  journal={bioRxiv},
  pages={2026--03},
  year={2026},
  publisher={Cold Spring Harbor Laboratory}
}

@article{baninoVectorbasedNavigationUsing2018,
  title = {Vector-Based Navigation Using Grid-like Representations in Artificial Agents},
  author = {Banino, Andrea and Barry, Caswell and Uria, Benigno and Blundell, Charles and Lillicrap, Timothy and Mirowski, Piotr and Pritzel, Alexander and Chadwick, Martin J. and Degris, Thomas and Modayil, Joseph and Wayne, Greg and Soyer, Hubert and Viola, Fabio and Zhang, Brian and Goroshin, Ross and Rabinowitz, Neil and Pascanu, Razvan and Beattie, Charlie and Petersen, Stig and Sadik, Amir and Gaffney, Stephen and King, Helen and Kavukcuoglu, Koray and Hassabis, Demis and Hadsell, Raia and Kumaran, Dharshan},
  year = 2018,
  month = may,
  journal = {Nature},
  volume = {557},
  number = {7705},
  pages = {429--433},
  issn = {0028-0836, 1476-4687},
  doi = {10.1038/s41586-018-0102-6},
  urldate = {2023-06-19},
  langid = {english}
}

@article{bennaPlaceCellsMay2021,
  title = {Place Cells May Simply Be Memory Cells: {{Memory}} Compression Leads to Spatial Tuning and History Dependence},
  shorttitle = {Place Cells May Simply Be Memory Cells},
  author = {Benna, Marcus K. and Fusi, Stefano},
  year = 2021,
  month = dec,
  journal = {Proc. Natl. Acad. Sci. U.S.A.},
  volume = {118},
  number = {51},
  pages = {e2018422118},
  issn = {0027-8424, 1091-6490},
  doi = {10.1073/pnas.2018422118},
  urldate = {2023-06-19},
  langid = {english}
}

@article{bonnevieGridCellsRequire2013,
  title = {Grid Cells Require Excitatory Drive from the Hippocampus},
  author = {Bonnevie, Tora and Dunn, Benjamin and Fyhn, Marianne and Hafting, Torkel and Derdikman, Dori and Kubie, John L and Roudi, Yasser and Moser, Edvard I and Moser, May-Britt},
  year = 2013,
  month = mar,
  journal = {Nat Neurosci},
  volume = {16},
  number = {3},
  pages = {309--317},
  issn = {1097-6256, 1546-1726},
  doi = {10.1038/nn.3311},
  urldate = {2025-02-10},
  langid = {english}
}

@article{burakAccuratePathIntegration2009,
  title = {Accurate {{Path Integration}} in {{Continuous Attractor Network Models}} of {{Grid Cells}}},
  author = {Burak, Yoram and Fiete, Ila R.},
  editor = {Sporns, Olaf},
  year = 2009,
  month = feb,
  journal = {PLoS Comput Biol},
  volume = {5},
  number = {2},
  pages = {e1000291},
  issn = {1553-7358},
  doi = {10.1371/journal.pcbi.1000291},
  urldate = {2025-02-10},
  langid = {english}
}

@article{burgessOscillatoryInterferenceModel2007,
  title = {An Oscillatory Interference Model of Grid Cell Firing},
  author = {Burgess, Neil and Barry, Caswell and O'Keefe, John},
  year = 2007,
  month = sep,
  journal = {Hippocampus},
  volume = {17},
  number = {9},
  pages = {801--812},
  issn = {1050-9631, 1098-1063},
  doi = {10.1002/hipo.20327},
  urldate = {2026-05-02},
  langid = {english}
}

@article{bushWhatGridCells2014,
  title = {What Do Grid Cells Contribute to Place Cell Firing?},
  author = {Bush, Daniel and Barry, Caswell and Burgess, Neil},
  year = 2014,
  month = mar,
  journal = {Trends in Neurosciences},
  volume = {37},
  number = {3},
  pages = {136--145},
  issn = {01662236},
  doi = {10.1016/j.tins.2013.12.003},
  urldate = {2026-03-27},
  langid = {english}
}

@article{carpenterGridCellsForm2015,
  title = {Grid {{Cells Form}} a {{Global Representation}} of {{Connected Environments}}},
  author = {Carpenter, Francis and Manson, Daniel and Jeffery, Kate and Burgess, Neil and Barry, Caswell},
  year = 2015,
  month = may,
  journal = {Current Biology},
  volume = {25},
  number = {9},
  pages = {1176--1182},
  issn = {09609822},
  doi = {10.1016/j.cub.2015.02.037},
  urldate = {2026-04-16},
  langid = {english}
}

@article{chandraEpisodicAssociativeMemory2025,
  title = {Episodic and Associative Memory from Spatial Scaffolds in the Hippocampus},
  author = {Chandra, Sarthak and Sharma, Sugandha and Chaudhuri, Rishidev and Fiete, Ila},
  year = 2025,
  month = feb,
  journal = {Nature},
  volume = {638},
  number = {8051},
  pages = {739--751},
  issn = {0028-0836, 1476-4687},
  doi = {10.1038/s41586-024-08392-y},
  urldate = {2026-05-02},
  langid = {english}
}

@inproceedings{chuUnfoldingBlackBox2025,
  title = {Unfolding the {{Black Box}} of {{Recurrent Neural Networks}} for {{Path Integration}}},
  booktitle = {{{NeurIPS}}},
  author = {Chu, Tianhao and Wu, Yuling and Burgess, Neil and Ji, Zilong and Wu, Si},
  year = 2025,
  langid = {english}
}

@misc{cuevaEmergenceGridlikeRepresentations2018,
  title = {Emergence of Grid-like Representations by Training Recurrent Neural Networks to Perform Spatial Localization},
  author = {Cueva, Christopher J. and Wei, Xue-Xin},
  year = 2018,
  month = mar,
  number = {arXiv:1803.07770},
  eprint = {1803.07770},
  primaryclass = {cs, q-bio, stat},
  publisher = {arXiv},
  urldate = {2023-06-19},
  archiveprefix = {arXiv},
  langid = {english}
}

@article{derdikmanFragmentationGridCell2009,
  title = {Fragmentation of Grid Cell Maps in a Multicompartment Environment},
  author = {Derdikman, Dori and Whitlock, Jonathan R and Tsao, Albert and Fyhn, Marianne and Hafting, Torkel and Moser, May-Britt and Moser, Edvard I},
  year = 2009,
  month = oct,
  journal = {Nat Neurosci},
  volume = {12},
  number = {10},
  pages = {1325--1332},
  issn = {1097-6256, 1546-1726},
  doi = {10.1038/nn.2396},
  urldate = {2026-04-16},
  langid = {english}
}

@article{keinathEnvironmentalDeformations2018,
  article_type = {journal},
  title = {Environmental deformations dynamically shift the grid cell spatial metric},
  author = {Keinath, Alexandra T and Epstein, Russell A and Balasubramanian, Vijay},
  editor = {Colgin, Laura and Frank, Michael J},
  volume = 7,
  year = 2018,
  month = {oct},
  pub_date = {2018-10-22},
  pages = {e38169},
  citation = {eLife 2018;7:e38169},
  doi = {10.7554/eLife.38169},
  url = {https://doi.org/10.7554/eLife.38169},
  journal = {eLife},
  issn = {2050-084X},
  publisher = {eLife Sciences Publications, Ltd},
}

@article{hardcastle2015environmental,
  title={Environmental boundaries as an error correction mechanism for grid cells},
  author={Hardcastle, Kiah and Ganguli, Surya and Giocomo, Lisa M},
  journal={Neuron},
  volume={86},
  number={3},
  pages={827--839},
  year={2015},
  publisher={Elsevier}
}

@article{sargolini2006conjunctive,
  title={Conjunctive representation of position, direction, and velocity in entorhinal cortex},
  author={Sargolini, Francesca and Fyhn, Marianne and Hafting, Torkel and McNaughton, Bruce L and Witter, Menno P and Moser, May-Britt and Moser, Edvard I},
  journal={Science},
  volume={312},
  number={5774},
  pages={758--762},
  year={2006},
  publisher={American Association for the Advancement of Science}
}

@article{stensolaShearinginducedAsymmetryEntorhinal2015,
  title = {Shearing-Induced Asymmetry in Entorhinal Grid Cells},
  author = {Stensola, Tor and Stensola, Hanne and Moser, May-Britt and Moser, Edvard I.},
  year = 2015,
  month = feb,
  journal = {Nature},
  volume = {518},
  number = {7538},
  pages = {207--212},
  issn = {0028-0836, 1476-4687},
  doi = {10.1038/nature14151},
  urldate = {2026-05-06},
  langid = {english}
}

@article{bushUsingGridCells2015,
  title = {Using {{Grid Cells}} for {{Navigation}}},
  author = {Bush, Daniel and Barry, Caswell and Manson, Daniel and Burgess, Neil},
  year = 2015,
  month = aug,
  journal = {Neuron},
  volume = {87},
  number = {3},
  pages = {507--520},
  issn = {08966273},
  doi = {10.1016/j.neuron.2015.07.006},
  urldate = {2025-01-29},
  langid = {english}
}

@article{krupicGridCellSymmetry2015,
  title = {Grid Cell Symmetry Is Shaped by Environmental Geometry},
  author = {Krupic, Julija and Bauza, Marius and Burton, Stephen and Barry, Caswell and O'Keefe, John},
  year = 2015,
  month = feb,
  journal = {Nature},
  volume = {518},
  number = {7538},
  pages = {232--235},
  issn = {0028-0836, 1476-4687},
  doi = {10.1038/nature14153},
  urldate = {2025-02-10},
  langid = {english}
}

@article{ginosarLocallyOrderedRepresentation2021,
  title = {Locally Ordered Representation of {{3D}} Space in the Entorhinal Cortex},
  author = {Ginosar, Gily and Aljadeff, Johnatan and Burak, Yoram and Sompolinsky, Haim and Las, Liora and Ulanovsky, Nachum},
  year = 2021,
  month = aug,
  journal = {Nature},
  volume = {596},
  number = {7872},
  pages = {404--409},
  issn = {0028-0836, 1476-4687},
  doi = {10.1038/s41586-021-03783-x},
  urldate = {2025-05-14},
  langid = {english}
}

@article{haftingMicrostructureSpatialMap2005,
  title = {Microstructure of a Spatial Map in the Entorhinal Cortex},
  author = {Hafting, Torkel and Fyhn, Marianne and Molden, Sturla and Moser, May-Britt and Moser, Edvard I.},
  year = 2005,
  month = aug,
  journal = {Nature},
  volume = {436},
  number = {7052},
  pages = {801--806},
  issn = {0028-0836, 1476-4687},
  doi = {10.1038/nature03721},
  urldate = {2024-09-18},
  langid = {english}
}

@article{kangGeometricAttractorMechanism2019,
  title = {A Geometric Attractor Mechanism for Self-Organization of Entorhinal Grid Modules},
  author = {Kang, Louis and Balasubramanian, Vijay},
  year = 2019,
  month = aug,
  journal = {eLife},
  volume = {8},
  doi = {10.7554/eLife.46687},
  langid = {english}
}

@article{khonaGlobalModulesRobustly2025,
  title = {Global Modules Robustly Emerge from Local Interactions and Smooth Gradients},
  author = {Khona, Mikail and Chandra, Sarthak and Fiete, Ila},
  year = 2025,
  month = apr,
  journal = {Nature},
  volume = {640},
  number = {8057},
  pages = {155--164},
  issn = {0028-0836, 1476-4687},
  doi = {10.1038/s41586-024-08541-3},
  urldate = {2026-04-30},
  langid = {english}
}

@article{langstonDevelopmentSpatialRepresentation2010,
  title = {Development of the {{Spatial Representation System}} in the {{Rat}}},
  author = {Langston, Rosamund F. and Ainge, James A. and Couey, Jonathan J. and Canto, Cathrin B. and Bjerknes, Tale L. and Witter, Menno P. and Moser, Edvard I. and Moser, May-Britt},
  year = 2010,
  month = jun,
  journal = {Science},
  volume = {328},
  number = {5985},
  pages = {1576--1580},
  issn = {0036-8075, 1095-9203},
  doi = {10.1126/science.1188210},
  urldate = {2025-05-06},
  langid = {english}
}

@article{morrisChickenEggProblem2023,
  title = {The Chicken and Egg Problem of Grid Cells and Place Cells},
  author = {Morris, Genela and Derdikman, Dori},
  year = 2023,
  month = feb,
  journal = {Trends in Cognitive Sciences},
  volume = {27},
  number = {2},
  pages = {125--138},
  issn = {13646613},
  doi = {10.1016/j.tics.2022.11.003},
  urldate = {2023-06-19},
  langid = {english}
}

@article{moserPlaceCellsGrid2008,
  title = {Place {{Cells}}, {{Grid Cells}}, and the {{Brain}}'s {{Spatial Representation System}}},
  author = {Moser, Edvard I. and Kropff, Emilio and Moser, May-Britt},
  year = 2008,
  month = jul,
  journal = {Annu. Rev. Neurosci.},
  volume = {31},
  number = {1},
  pages = {69--89},
  issn = {0147-006X, 1545-4126},
  doi = {10.1146/annurev.neuro.31.061307.090723},
  urldate = {2023-09-22},
  langid = {english}
}

@article{moserPlaceCellsGrid2015,
  title = {Place {{Cells}}, {{Grid Cells}}, and {{Memory}}},
  author = {Moser, May-Britt and Rowland, David C. and Moser, Edvard I.},
  year = 2015,
  month = feb,
  journal = {Cold Spring Harb Perspect Biol},
  volume = {7},
  number = {2},
  pages = {a021808},
  issn = {1943-0264},
  doi = {10.1101/cshperspect.a021808},
  urldate = {2023-11-04},
  langid = {english}
}

@article{muessigDevelopmentalSwitchPlace2015,
  title = {A {{Developmental Switch}} in {{Place Cell Accuracy Coincides}} with {{Grid Cell Maturation}}},
  author = {Muessig, Laurenz and Hauser, Jonas and Wills, Thomas~Joseph and Cacucci, Francesca},
  year = 2015,
  month = jun,
  journal = {Neuron},
  volume = {86},
  number = {5},
  pages = {1167--1173},
  issn = {08966273},
  doi = {10.1016/j.neuron.2015.05.011},
  urldate = {2025-05-06},
  langid = {english}
}

@article{okeefeHippocampusSpatialMap1971,
  title = {The Hippocampus as a Spatial Map. {{Preliminary}} Evidence from Unit Activity in the Freely-Moving Rat},
  author = {O'Keefe, J. and Dostrovsky, J.},
  year = 1971,
  month = nov,
  journal = {Brain Research},
  volume = {34},
  number = {1},
  pages = {171--175},
  issn = {00068993},
  doi = {10.1016/0006-8993(71)90358-1},
  urldate = {2024-04-17},
  langid = {english}
}

@article{okeefePlaceUnitsHippocampus1976,
  title = {Place Units in the Hippocampus of the Freely Moving Rat},
  author = {O'Keefe, John},
  year = 1976,
  month = jan,
  journal = {Experimental Neurology},
  volume = {51},
  number = {1},
  pages = {78--109},
  issn = {00144886},
  doi = {10.1016/0014-4886(76)90055-8},
  urldate = {2026-03-27},
  langid = {english}
}

@article{rennó-costaPlaceGridCells2017,
  title = {Place and {{Grid Cells}} in a {{Loop}}: {{Implications}} for {{Memory Function}} and {{Spatial Coding}}},
  shorttitle = {Place and {{Grid Cells}} in a {{Loop}}},
  author = {{Renn{\'o}-Costa}, C{\'e}sar and Tort, Adriano B.L.},
  year = 2017,
  month = aug,
  journal = {J. Neurosci.},
  volume = {37},
  number = {34},
  pages = {8062--8076},
  issn = {0270-6474, 1529-2401},
  doi = {10.1523/JNEUROSCI.3490-16.2017},
  urldate = {2026-03-27},
  langid = {english}
}

@article{rollsEntorhinalCortexGrid2006,
  title = {Entorhinal Cortex Grid Cells Can Map to Hippocampal Place Cells by Competitive Learning},
  author = {Rolls, Edmund T. and Stringer, Simon M. and Elliot, Thomas},
  year = 2006,
  month = jan,
  journal = {Network: Computation in Neural Systems},
  volume = {17},
  number = {4},
  pages = {447--465},
  issn = {0954-898X, 1361-6536},
  doi = {10.1080/09548980601064846},
  urldate = {2026-03-27},
  langid = {english}
}

@inproceedings{schaefferSelfSupervisedLearningRepresentations2023,
  title = {Self-{{Supervised Learning}} of {{Representations}} for {{Space Generates Multi-Modular Grid Cells}}},
  booktitle = {{{NeurIPS}}},
  author = {Schaeffer, Rylan and Ma, Tzuhsuan and Koyejo, Sanmi and Khona, Mikail and Eyzaguirre, Crist{\'o}bal and Fiete, Ila Rani},
  year = 2023,
  month = sep,
  langid = {english}
}

@article{solstadGridCellsPlace2006,
  title = {From Grid Cells to Place Cells: {{A}} Mathematical Model},
  shorttitle = {From Grid Cells to Place Cells},
  author = {Solstad, Trygve and Moser, Edvard I. and Einevoll, Gaute T.},
  year = 2006,
  month = dec,
  journal = {Hippocampus},
  volume = {16},
  number = {12},
  pages = {1026--1031},
  issn = {1050-9631, 1098-1063},
  doi = {10.1002/hipo.20244},
  urldate = {2026-03-27},
  langid = {english}
}

@article{sorscherUnifiedTheoryComputational2023,
  title = {A Unified Theory for the Computational and Mechanistic Origins of Grid Cells},
  author = {Sorscher, Ben and Mel, Gabriel C. and Ocko, Samuel A. and Giocomo, Lisa M. and Ganguli, Surya},
  year = 2023,
  month = jan,
  journal = {Neuron},
  volume = {111},
  number = {1},
  pages = {121-137.e13},
  issn = {08966273},
  doi = {10.1016/j.neuron.2022.10.003},
  urldate = {2023-06-19},
  langid = {english}
}

@article{stachenfeldHippocampusPredictiveMap2017,
  title = {The Hippocampus as a Predictive Map},
  author = {Stachenfeld, Kimberly L and Botvinick, Matthew M and Gershman, Samuel J},
  year = 2017,
  month = nov,
  journal = {Nat Neurosci},
  volume = {20},
  number = {11},
  pages = {1643--1653},
  issn = {1097-6256, 1546-1726},
  doi = {10.1038/nn.4650},
  urldate = {2023-12-07},
  langid = {english}
}

@article{tsodyksParadoxicalEffectsExternal1997,
  title = {Paradoxical {{Effects}} of {{External Modulation}} of {{Inhibitory Interneurons}}},
  author = {Tsodyks, Misha V. and Skaggs, William E. and Sejnowski, Terrence J. and McNaughton, Bruce L.},
  year = 1997,
  month = jun,
  journal = {J. Neurosci.},
  volume = {17},
  number = {11},
  pages = {4382--4388},
  issn = {0270-6474, 1529-2401},
  doi = {10.1523/JNEUROSCI.17-11-04382.1997},
  urldate = {2026-04-29},
  langid = {english}
}

@inproceedings{wangREMIReconstructingEpisodic2025,
  title = {{{REMI}}: {{Reconstructing Episodic Memory During Internally Driven Path Planning}}},
  booktitle = {{{NeurIPS}}},
  author = {Wang, Zhaoze and Morris, Genela and Derdikman, Dori and Chaudhari, Pratik and Balasubramanian, Vijay},
  year = 2025,
  langid = {english}
}

@inproceedings{wangTimeMakesSpace2024,
  title = {Time {{Makes Space}}: {{Emergence}} of {{Place Fields}} in {{Networks Encoding Temporally Continuous Sensory Experiences}}},
  shorttitle = {Time {{Makes Space}}},
  booktitle = {{{NeurIPS}}},
  author = {Wang, Zhaoze and Di Tullio, Ronald W. and Rooke, Spencer and Balasubramanian, Vijay},
  year = 2024,
  doi = {10.1101/2024.08.11.607484},
  langid = {english}
}

@article{willsDevelopmentHippocampalCognitive2010,
  title = {Development of the {{Hippocampal Cognitive Map}} in {{Preweanling Rats}}},
  author = {Wills, Tom J. and Cacucci, Francesca and Burgess, Neil and O'Keefe, John},
  year = 2010,
  month = jun,
  journal = {Science},
  volume = {328},
  number = {5985},
  pages = {1573--1576},
  issn = {0036-8075, 1095-9203},
  doi = {10.1126/science.1188224},
  urldate = {2025-05-06},
  langid = {english}
}

@misc{xuConformalIsometryGrid2025,
  title = {On {{Conformal Isometry}} of {{Grid Cells}}: {{Learning Distance-Preserving Position Embedding}}},
  shorttitle = {On {{Conformal Isometry}} of {{Grid Cells}}},
  author = {Xu, Dehong and Gao, Ruiqi and Zhang, Wen-Hao and Wei, Xue-Xin and Wu, Ying Nian},
  year = 2025,
  month = feb,
  number = {arXiv:2405.16865},
  eprint = {2405.16865},
  primaryclass = {q-bio},
  publisher = {arXiv},
  doi = {10.48550/arXiv.2405.16865},
  urldate = {2025-04-26},
  archiveprefix = {arXiv},
  langid = {english}
}

@article{vanVreeswijk1996,
  author  = {van Vreeswijk, Carl and Sompolinsky, Haim},
  title   = {Chaos in neuronal networks with balanced excitatory and inhibitory activity},
  journal = {Science},
  year    = {1996},
  volume  = {274},
  number  = {5293},
  pages   = {1724--1726},
  note    = {Accessed: 29 Apr 2026}
}


\appendix

\newpage

\renewcommand{\thesection}{S}
\renewcommand{\thesubsection}{S.\arabic{subsection}}
\renewcommand{\thesubsubsection}{S.\arabic{subsection}.\arabic{subsubsection}}
\setcounter{section}{0}
\setcounter{subsection}{0}
\setcounter{figure}{0}
\setcounter{table}{0}
\setcounter{equation}{0}
\renewcommand{\thefigure}{S\arabic{figure}}
\renewcommand{\thetable}{S\arabic{table}}
\renewcommand{\theequation}{S\arabic{equation}}

\section{Supplementary material}


\subsection{Continuous-time RNN dynamics}
\label{sup:continuous_rnn}

The discrete-time update in Eq.~\ref{eq:rnn_discrete} is derived from the following continuous-time neural dynamics:
\begin{equation}
\tau \frac{d h_i(t)}{dt}
=
- h_i(t)
+ \sum_{j=1}^{N} W_{ij}\, \phi\!\big(h_j(t)\big)
+ \sum_{k=1}^{d_{\text{in}}} B_{ik}\, u_k(t)
+ b_i,
\label{eq:rnn_continuous}
\end{equation}
where \(\tau\) is the membrane time constant. Applying Euler discretization with step \(\Delta t\) and defining the leak rate \(\alpha = \Delta t / \tau\), we obtain the update rule used in the main text. Throughout this work we use \(\phi(\cdot) = \mathrm{softplus}(\cdot)\) as the firing-rate nonlinearity, which provides a smooth non-negative activation. Biases \(b_i\) are fixed at zero.

\subsection{Excitatory-inhibitory structure and homeostatic synaptic scaling}
\label{sup:dale}

\textbf{Sign assignment.} The recurrent matrix \(W \in \mathbb{R}^{N \times N}\) acts on a hidden vector of size \(N = N_{\text{input-drv}} + N_{\text{rec-drv}}\), with \(N_{\text{input-drv}} = N_{\text{rec-drv}} = 1024\) in all main experiments. Each column \(j\) of \(W\) (the outgoing weights of neuron \(j\)) is constrained to a single sign: \(80\%\) of neurons are randomly assigned to be excitatory (sign \(+1\)) and \(20\%\) inhibitory (sign \(-1\)). The sign vector \(s \in \{-1, +1\}^N\) is drawn at initialization and held fixed throughout the training. We parameterize the recurrent matrix as
\begin{equation}
W_{ij} = |M_{ij}|\, s_j,
\label{eq:dale_param}
\end{equation}
where \(M \in \mathbb{R}^{N \times N}\) is a learnable magnitude matrix. Because only \(|M|\) enters Eq.~\ref{eq:dale_param}, the sign of \(M_{ij}\) is irrelevant and the column sign \(s_j\) is preserved through every gradient step, enforcing Dale's law without projection.

\textbf{Initialization.} The magnitude matrix \(M\) is initialized as in Algorithm~\ref{alg:init}: sample a Gaussian matrix, apply Dale signs columnwise, normalize the spectral radius to \(1\), and store the magnitudes into \(M\). For each neuron \(j\), we record its total outgoing synaptic strength \(c^{*}_j = \sum_i M_{ij}\) for the homeostatic scaling (Eq.~\ref{eq:homeostasis}). We describe the motivation for this homeostatic constraint and its role during training in the next section.

\begin{algorithm}[H]
\caption{Recurrent weight initialization}
\label{alg:init}
\begin{algorithmic}[1]
\State Sample \(\tilde{W}_{ij} \sim \mathcal{N}(0, 1/N)\)
\State \(\tilde{W}_{ij} \leftarrow s_j \, |\tilde{W}_{ij}|\) \Comment{apply Dale signs}
\State \(\tilde{W} \leftarrow \tilde{W} / \rho(\tilde{W})\) \Comment{normalize spectral radius to 1}
\State \(M_{ij} \leftarrow |\tilde{W}_{ij}|\) \Comment{store magnitudes; signs reapplied via Eq.~\ref{eq:dale_param}}
\State \(c^{*}_j \leftarrow \sum_i M_{ij}\) \Comment{total outgoing synaptic strength of neuron \(j\)}
\end{algorithmic}
\end{algorithm}

\textbf{Homeostatic synaptic scaling.} Back-propagation in the RNN may update synaptic weights much faster than biological plasticity would allow, which can destabilize the training dynamics. To mitigate this, after each weight update, we apply a normalization that softly pulls each neuron's total outgoing synaptic strength toward its initial value:
\begin{equation}
M_{ij} \leftarrow M_{ij} \left( \frac{c^{*}_j}{\sum_k M_{kj}} \right)^{\eta},
\label{eq:homeostasis}
\end{equation}
with \(\eta = 10^{-3}\). This regularization does not prevent co-emergence, but without it the network often fails to learn new environments and instead overfits to the first room.

\textbf{Multiplicative firing-rate noise.} Previous studies have suggested that place cells may emerge from denoising corrupted sensory input \cite{wangTimeMakesSpace2024}. We consider two sources of such corruption. The first kind is partial occlusion of sensory cues during navigation. We model this by randomly masking sensory input channels independently across time and sensory dimensions, replacing masked values with the mean value of the corresponding sensory channel. The second is stochasticity in neural activity, this models cortical firing which is approximately Poisson-like with rate-dependent variance. Since our model operates on firing rates rather than explicit spikes, we approximate this variability by adding Gaussian noise whose magnitude depends on the firing rate:
\begin{equation}
\tilde{h}_i \leftarrow \max\!\left( 0,\ h_i + \sqrt{\sigma_n h_i + \epsilon}\, \xi_i \right), \quad \xi_i \sim \mathcal{N}(0, 1),
\label{eq:fr_noise}
\end{equation}
where \(\sigma_n\) is the noise level. Clipping enforces non-negativity of firing rates in our model, since actual spike counts are strictly non-negative.

\textbf{Input and output projections.} In our network, we split the hidden layer into two regions: (i) where neurons are directly driven by the sensory and motion input signal to predict the next sensory observation, and (ii) where neurons receive input only indirectly from the first region. We refer to the first region as the sensory-driven region, and the second as recurrent-driven region. The sensory-driven population receives input \(u_t = [\tilde{o}_t;\, m_t] \in \mathbb{R}^{d_{\text{in}}}\) with \(d_{\text{in}} = d_{\text{obs}} + d_{\text{motion}}\) through an input projection matrix \(B \in \mathbb{R}^{N_{\text{input-drv}} \times d_{\text{in}}}\). The free population receives no external input. The hidden state evolves according to Eq.~\ref{eq:rnn_discrete}:
\begin{equation}
h_{i,t+1} = (1 - \alpha)\, h_{i,t}
+ \alpha \Big(
\textstyle\sum_{j=1}^{N} W_{ij}\,\phi(h_{j,t})
+
\sum_{k=1}^{d_{\text{in}}} B_{ik}\, u_{k,t}
+
b_i
\Big),
\label{eq:rnn_discrete_sup}
\end{equation}
After this deterministic recurrent update, we apply the firing-rate-dependent noise in Eq.~\ref{eq:fr_noise} to each hidden unit. Predictions are read out from the sensory-driven population using a linear readout projection matrix: \(\hat{o}_t = C\, h^{\text{sens-drv}}_t\). The network is trained to minimize the error of predicting the next observation: \(\sum_t \| \hat{o}_t - o_{t+1} \|_2^2\).

\subsection{Simulating sensory observations}
\label{sup:sensory}

\textbf{Boundary-aware sensory input.} We simulate sensory observations using a method similar to \cite{wangTimeMakesSpace2024}. As they suggest, animals receive sensory experiences that change smoothly with location. To model this, we use random Gaussian fields for each sensory channel. Along each sensory channel, the signal varies smoothly across spatial locations, generating a smooth sensory response map of dimension \(W \times H\), where \(W\) and \(H\) are the dimensions of the environment in pixels, respectively. We suppose that an animal has multiple such smooth spatial channels; together, these channels construct a sensory response map of dimension \(d_{\text{sens}} \times W \times H\), where the sensory input at each location is defined by its location-specific sensory vector.
The authors of \cite{wangTimeMakesSpace2024} first initialize each sensory channel's spatial response map as a random Gaussian field, and then convolve the field with a 2D Gaussian smoothing kernel where the kernel width is \(\sigma\). However, this method does not respect environmental boundaries. Near a boundary, if \(\sigma\) is larger than the spatial width of a wall, then the 2D Gaussian smoothing kernel can blend sensory observations across the wall.

To resolve the above issue, we replace Gaussian smoothing with a boundary-aware diffusion process that runs for \(n_{\text{iter}}\) steps. At each step, we apply a \(3 \times 3\) box filter at each location that averages the sensory responses at neighboring locations that lie in free space. This prevents smoothing across walls and respects environmental boundaries. We set \(n_{\text{iter}} = \lceil 1.5 \sigma^2 \rceil\) because since one step of a \(3 \times 3\) box filter has variance \(2/3\) along each spatial dimension; so after \(n\) steps the effective Gaussian width is \(\sqrt{2n/3}\). This gives \(n \approx 1.5\sigma^2\) for a smoothing width \(\sigma\). 

\textbf{Populations.} We use four sensory input populations with different smoothing widths \(\sigma \in \{6, 8, 10, 12\}\) cm, each containing \(N_{\text{cell}} = 256\) cells, giving a total sensory dimension of \(d_{\text{obs}} = 1024\). Using multiple smoothing widths allows the sensory observations to contain both fine- and coarse-scale spatial features.

\subsection{Trajectory generation}
\label{sup:trajectory}

\textbf{Trajectory generation.} We simulate animal trajectories as a smooth random walk. At each timestep \(t\), the agent maintains a current speed \(v_t\) and unit movement direction \(\mathbf{d}_t\), together with a target speed \(v_t^{*}\) and target direction \(\mathbf{d}_t^{*}\) unit vector.
The target variables represent the speed and direction that the agent is currently trying to move toward. Instead of choosing a completely new velocity at every timestep, the agent only occasionally resamples these targets: \(v_t^{*}\) is redrawn with probability \(p_v\) and \(\mathbf{d}_t^{*}\) is redrawn with probability \(p_d\). This way, the agent tends to continue moving in a similar way for several timesteps before changing its intended motion. Speed targets are sampled from a log-normal distribution with mean \(\mu_{\text{spd}}\) and standard deviation \(\sigma_{\text{spd}}\) (speed is positive). Directions are sampled as random unit vectors in \(\mathbb{R}^2\) or \(\mathbb{R}^3\) depending on the simulation environment. To avoid abrupt changes when a new target is sampled, the current speed and direction are gradually updated toward their targets using exponential moving averages: \(v_{t+1} = (1-\alpha_v)v_t + \alpha_v v_t^{*}\), and \(\mathbf{d}_{t+1} = \mathrm{normalize}((1-\alpha_d)\mathbf{d}_t + \alpha_d \mathbf{d}_t^{*})\). Here, \(\alpha_v\) and \(\alpha_d\) control how quickly the simulated agent adapts to the newly sampled target speed and direction.

\textbf{Boundary avoidance.} To prevent the agent from running into walls or getting stuck near boundaries, we apply a soft boundary avoidance near walls. Before trajectory generation, we compute a distance-to-wall map \(d(x)\), which gives the distance from each location \(x\) to the nearest wall and a local wall-normal direction \(\mathbf{n}(x)\) which points away from the wall. We also define an avoidance strength \(c(x)\) which is large near walls and decays as the agent moves farther away from them. When the agent is far from any wall, its movement direction \(\mathbf{d}_t\) is unchanged. When the agent is close to a wall and moving towards it, we compute the wall-parallel direction by removing the component of \(\mathbf{d}_t\) that points into the wall. We then blend the current direction with this wall-parallel direction, with a blending weight \(\beta_t\) that increases when the agent is closer to the wall and when its speed is larger. This softly turns the agent along the boundary instead of abruptly reflecting it, producing smooth wall-following trajectories in both 2D and 3D environments.

\textbf{Training along long trajectories.} Previous RNN studies \cite{wangTimeMakesSpace2024, sorscherUnifiedTheoryComputational2023, cuevaEmergenceGridlikeRepresentations2018} train their models with backpropagation through time (BPTT) on short trajectory segments, typically around 20--40 simulation steps or approximately 1--2 s. In their setup, the network is updated using inputs sampled from short trajectory fragments; the simulated agent is relocated to a new position after each update. The authors in \cite{sorscherUnifiedTheoryComputational2023} initialize the network hidden state with a small multilayer perceptron after each location reset, while the authors in \cite{wangTimeMakesSpace2024} initialize the hidden state to zero. Our model is also trained with BPTT. But we realized that such reset-based training schemes do not allow studying place-cell and grid-cell behavior over longer continuous trajectories. This is especially true in experiments such as the hairpin maze. We resolve this issue by detaching the last hidden state from the previous training segment and use it to initialize the first hidden state of the current training segment. This is the appropriate way to implement truncated BPTT and it allows the model to be trained on short segments while preserving continuous trajectory dynamics across updates.

\subsection{Sensory masking}
\label{sup:masking}

As discussed in Section~\ref{sup:continuous_rnn}, we simulate noisy sensory inputs by masking parts of it by the mean sensory response.
Masking is applied independently across the batch, timesteps and sensory dimensions.
At each timestep, each sensory dimension is independently masked with probability \(r_{\text{mask}}\), where \(r_{\text{mask}}\) denotes the masking ratio. Masked entries are replaced by the corresponding cell's mean firing rate, computed over the arena, while unmasked entries keep their original sensory value. Using the per-cell mean rather than zero prevents the network from treating masked entries as an artificial zero-valued cue.

\subsection{Collecting ratemaps during training}
\label{sup:ratemaps}

Experimentally measured ratemaps are obtained while the animal explores actively. Thus, instead of freezing the model and evaluating it in a separate testing phase, we continuously compute the firing rate of different cells using the trajectories sampled during training.
At each optimization step, we collect hidden states from all batches and timesteps, together with the corresponding agent coordinates. The coordinates are rounded to the nearest spatial bin. For each visited bin, we compute each hidden unit's mean firing rate over all visits to that bin. This gives a ratemap estimate over the subset of spatial bins visited during the current optimization step.
To aggregate ratemaps over training, we maintain one running ratemap for each hidden unit. At each update step, only the spatial bins visited by the current trajectories are updated. For those bins, the ratemap values from previous updates are combined with the average firing rates computed at the current update step using an exponential moving average with decay \(\gamma = 0.995\). Spatial bins that are not visited keep their previous values and bins that have never been visited are marked as undefined. This allows ratemaps to be accumulated smoothly throughout training while giving slightly higher weight to more recent activity. With \(\gamma = 0.995\), the ratemap retains an effective memory of approximately 200 update steps for regularly visited bins.

\subsection{Default training parameters}
\label{sup:parameters}

Unless otherwise stated, all reported runs use the parameters in Table~\ref{tab:default_params}. Per-experiment overrides are listed in their respective sections.

\begin{table}
\centering
\caption{Default training and architectural parameters.}
\label{tab:default_params}
\renewcommand{\arraystretch}{1.2}
\begin{tabular}{p{0.25\linewidth} p{0.2\linewidth} p{0.4\linewidth}}
\toprule
\textbf{Parameter} & \textbf{Value} & \textbf{Description} \\
\midrule
\(N_{\text{input-drv}}\) & 1024 & Sensory-driven population size \\
\(N_{\text{rec-drv}}\) & 1024 & Free population size \\
\(\phi(\cdot)\) & softplus & Firing-rate nonlinearity \\
EI ratio & 0.8 / 0.2 & Excitatory / inhibitory fraction \\
\(\alpha\) & 0.5 & Default leak rate (sigmoid-gated) \\
\(\sigma_n\) & 0.8 & Default firing-rate noise level \\
\(b_i\) & 0 (fixed) & Bias term in Eq.~\ref{eq:rnn_discrete} \\
\(\eta\) & \(10^{-3}\) & Homeostatic scaling exponent \\
optimizer & AdamW & \\
learning\_rate & \(5 \cdot 10^{-4}\) & \\
batch\_size & 256 & Each batch is one batch of trajectories \\
traj\_duration & 10 & Backpropagation window \\
gradient\_clip & 1.0 & L2 norm cap on parameter gradients \\
\(\Delta t\) & 1/20 s & Simulation timestep \\
\(\mu_{\text{spd}}, \sigma_{\text{spd}}\) & 40, 20 cm/s & Log-normal speed mean and std \\
\(\alpha_v, \alpha_d\) & 0.8, 0.2 & EMA smoothing for speed and direction \\
\(p_v, p_d\) & 0.2, 0.1 & Per-step target switch probabilities \\
\(\sigma_{\text{cell}}\) & \(\{6, 8, 10, 12\}\) & Sensory cells smoothing widths \\
\(N_{\text{cell}}\) & 256 per population & Cells per sensory population \\
\(r_{\text{mask}}\) & 0.2 & Default sensory mask ratio \\
\bottomrule
\end{tabular}
\end{table}


\subsection{Cell classification}
\label{sup:cell_classification}

We classify cells into three types: \emph{place}, \emph{grid} or \emph{other spatial}, based on per-cell scores computed from each ratemap.

\textbf{Classifying activated spatial cells.} Since both grid cells and place cells are strongly modulated by an animal's spatial locations, we first filter out cells that are active and have spatial information content above 0.2. We define a cell as active if it has a mean firing rate \(\bar{r}_i \le 0.1\). The spatial information content is
\begin{equation}
\mathrm{SIC}_i =
\sum_{\mathbf{x}} p(\mathbf{x})
\frac{R_i(\mathbf{x})}{\bar{r}_i}
\log_2 \frac{R_i(\mathbf{x})}{\bar{r}_i},
\label{eq:sic}
\end{equation}
where \(p(\mathbf{x})\) is the empirical occupancy of spatial bin \(\mathbf{x}\). Cells with \(\mathrm{SIC}_i > 0.2\) are treated as spatial cells.

\textbf{Place-cell classification.}
We next test whether selected spatial cells are place cells. Previous studies often use SIC alone to identify place cells. However, because our model contains place cells, grid cells, and other spatially tuned cells within the same hidden layer, SIC alone may not be sufficient for robust classification. We therefore add two firing-field locality criteria for place-cell classification. Specifically, a place cell should contain only a small number of spatially localized firing fields, rather than many repeated peaks. In addition, its firing should be concentrated to a small spatial region.
To measure this, we first threshold each ratemap at \(0.3\) of its maximum firing rate. After thresholding, each connected group of spatial bins above the threshold is identified as a firing field. We classify a cell as a place cell if it has fewer than \(3\) such firing fields and if its largest field contains at least \(35\%\) of the cell's total firing mass.

\def \btau {\boldsymbol{\tau}}
\def \bx {\mathbf{x}}

\textbf{Spatial autocorrelogram.} To classify grid cells, previous studies have used autocorrelograms to reveal the characteristic spatial periodicity of grid-cell firing \cite{haftingMicrostructureSpatialMap2005}. In this study, we use the same spatial autocorrelogram computation method as that of \cite{haftingMicrostructureSpatialMap2005}. Specifically, for each ratemap \(R_i(\mathbf{x})\), we compute a 2D spatial autocorrelogram to measure whether the firing fields form a regular periodic structure. For each spatial lag \(\boldsymbol{\tau} = (\tau_x,\tau_y)\), we compare the ratemap with a shifted copy of itself, \(R_i(\mathbf{x}-\boldsymbol{\tau})\). The correlation is computed only over spatial bins where both \(R_i(\mathbf{x})\) and \(R_i(\mathbf{x}-\boldsymbol{\tau})\) are valid, which corrects for edge effects and unvisited locations.
Following this standard sample-correlation formulation, the autocorrelation at lag
\(\boldsymbol{\tau} = (\tau_x,\tau_y)\) for cell \(i\), with ratemap \(R \equiv R_i\), is
\begin{equation}
A(\boldsymbol{\tau}) =
\frac{
n\sum R(\mathbf{x}) R(\mathbf{x}-\boldsymbol{\tau})
-
\sum R(\mathbf{x})\sum R(\mathbf{x}-\boldsymbol{\tau})
}{
\sqrt{
n\sum R(\mathbf{x})^2-\left(\sum R(\mathbf{x})\right)^2
}
\sqrt{
n\sum R(\mathbf{x}-\boldsymbol{\tau})^2-\left(\sum R(\mathbf{x}-\boldsymbol{\tau})\right)^2
}
},
\label{eq:autocorr}
\end{equation}
where all sums are taken over the \(n\) spatial bins for which both
\(R(\mathbf{x})\) and \(R(\mathbf{x}-\boldsymbol{\tau})\) are valid. Lags with too few overlapping valid bins are not evaluated. The resulting autocorrelogram is shifted so that the zero-lag term is centered, producing a \((2H-1)\times(2W-1)\) map for an \(H\times W\) ratemap.

\textbf{Grid score.} The ``grid score'' measures whether a cell has the six-fold rotational symmetry characteristic of grid cells. Following the standard grid-score procedure \cite{sargolini2006conjunctive}, we take a centered crop of the autocorrelogram computed above with radius \(r=60\) bins. We exclude the small central peak around zero lag, since this peak appears for all cells, and also exclude a narrow edge band to avoid boundary artifacts.
We then compute the Pearson correlation between the autocorrelogram and a rotated copy of itself. Let \(\rho_{\theta}\) denote this correlation after rotation by \(\theta\) degrees. The grid score is defined as
\begin{equation}
\mathrm{GS}_i =
\min(\rho_{60}, \rho_{120})
-
\max(\rho_{30}, \rho_{90}, \rho_{150}).
\label{eq:grid_score}
\end{equation}
A hexagonal grid should have high correlation after \(60^\circ\) and \(120^\circ\) rotations, but lower correlation after \(30^\circ\), \(90^\circ\), and \(150^\circ\) rotations.

Note that we classify grid cells after place-cell classification. This ordering is unlikely to remove valid grid cells, because our place-cell criterion requires fewer than \(3\) firing fields, whereas a valid grid cell should contain multiple periodically repeated fields. Among the remaining spatially tuned cells, we classify a cell as a grid cell if it passes the grid-score threshold. We use \(\mathrm{GS}_i > 0.3\) for all analyses, except for the parameter-sweep result in Figure~\ref{fig:main_results}F. For the parameter sweep, we use a more lenient threshold of \(\mathrm{GS}_i > 0.1\), so that the summary better reflects how grid-like periodic patterns change across the swept parameters. We also require \(\min(\rho_{60},\rho_{120})>0\), ensuring that a positive grid score is supported by positive correlations at the grid-symmetric rotations. Finally, we require at least \(4\) firing fields, which prevents single-peaked or weakly localized spatial cells from being classified as grid cells.

\textbf{Other spatial cells.} Spatial cells that do not satisfy either the place or grid criteria are labelled \emph{other spatial}. These are cells with structured but non-localized and non-hexagonal spatial tuning (e.g., stripe-like, multi-peaked irregular fields, or border-like).

\subsection{Parameter sweep details}
\label{sup:parameter_sweep}

\textbf{Sweep grid.} The parameter sweep in the main text (Figure~\ref{fig:main_results}F) covers a \(10 \times 10 \times 10\) grid over the leak rate, firing-rate noise level and sensory mask ratio:
\begin{equation}
\alpha \in \{0.1, 0.2, \ldots, 1.0\}, \quad
\sigma_n \in \{0, 0.05, \ldots, 0.45\}, \quad
r_{\text{mask}} \in \{0, 0.1, \ldots, 0.9\}.
\end{equation}
For each configuration, we train the network for 200,000 weight updates in the \(220\,\text{cm} \times 220\,\text{cm}\) square arena while holding all other parameters at their default values (Table~\ref{tab:default_params}). We use a different random seed for each training condition to test robustness across stochastic components of the setup, including the sensory response fields, trajectories, and network initialization. We use only one random seed for each configuration (different for each configuration).
When summarizing results along each marginal axis, each value aggregates over \(100\) sweep conditions, providing implicit averaging over the other parameters and random seeds.

\subsection{Two-room wall removal experiment}
\label{sup:two_rooms}

In this experiment, two \(180 \times 150\,\text{cm}^2\) rectangular rooms are placed side-by-side and separated by a thin internal wall of thickness equal to twice the simulator's \(5\)-cm border, giving a merged arena of size \(180 \times 310\,\text{cm}^2\). We train the network over three phases, each with \(100{,}000\) network updates:
\begin{enumerate}[nosep,leftmargin=*,topsep=2pt]
\item Room 0 (left sub-room). The agent is spawned and confined to the left compartment.
\item Room 1 (right sub-room). The agent is respawned in the right compartment.
\item Merged arena. The internal wall is removed, and the agent is allowed to move freely through the full arena.
\end{enumerate}

At initialization, we first sample a random Gaussian noise field. We then generate the sensory response maps for the two rooms using the boundary-aware diffusion smoothing method described in Section~\ref{sup:sensory}. Because this diffusion process respects environmental boundaries, sensory values are not smoothed across the wall separating the two rooms. As a result, the sensory response maps on the two sides of the internal wall can differ sharply. During the first two phases, the agent is spawned in the corresponding room, and its trajectory is constrained to be within that room because the two rooms are not connected.

After the agent has explored both rooms, we remove the internal wall. Starting from the same initial Gaussian noise field, we re-apply the diffusion-based smoothing process on the merged arena. After this re-smoothing, sensory responses near the original wall location become smoothly connected across the two formerly separated rooms. Since the diffusion process only acts locally, sensory responses at locations far from the removed wall are expected to remain similar to their values before wall removal.

Across all three phases, we use the parameters in Table~\ref{tab:default_params}, with \(\alpha = 0.5\), \(\sigma_n = 0.8\), and \(r_{\text{mask}} = 0\). We maintain a ratemap aggregator over the full merged arena. During phases 0 and 1, only the bins visited in the active room are updated, while bins in the other room remain unchanged. At the beginning of phase 2, we reset the ratemap aggregation so that the merged-arena ratemaps reflect only the post-wall-removal experience.

\subsection{Hairpin maze experiment}
\label{sup:hairpin}

The hairpin arena contains \(10\) parallel corridors connected by alternating turn gaps, forming a continuous serpentine path. Its bounding box is matched to the open arena, so the same raw sensory field can be evaluated in both environments after applying boundary-aware diffusion smoothing. To construct the sensory inputs, we first generate one raw Gaussian noise field for each sensory population. We then apply the boundary-aware diffusion process separately in the open arena and in the hairpin maze. Because the raw field is shared, corresponding locations have similar sensory values away from the inserted walls. In the hairpin maze, however, diffusion is blocked by corridor walls, so sensory similarity follows the corridor structure rather than passing through walls.

In the open-arena phases, trajectories are generated by the default smooth random-walk generator. In the hairpin phases, we use a scripted serpentine controller that drives the agent through the corridors one lap at a time: left-to-right in the second phase (Figure~\ref{fig:multi_compartment}B, MazeLR) and right-to-left in the third phase (Figure~\ref{fig:multi_compartment}B, MazeRL). The hidden state is reset to zero at each lap boundary, so successive laps form independent trials. At every step, speed is sampled from the same log-normal distribution used by the open-arena controller. We also add small random lateral offsets within each corridor, together with a perpendicular drift, so that the agent samples the full corridor width rather than only following the centerline.

Following \cite{derdikmanFragmentationGridCell2009}, we train the model across four phases: open arena, hairpin forward, hairpin backward, and open arena again. During each phase, we update the network for \(100,000\) steps. We reset the accumulated ratemap whenever we shift the training phase, so each phase's ratemaps reflect only cell activity collected during that phase.

\subsection{Connected-rooms experiment}
\label{sup:connected_rooms}

We use two \(120 \times 120\,\text{cm}^2\) compartments separated by an \(8\,\text{cm}\) dividing wall, with a \(248 \times 70\,\text{cm}^2\) return corridor along the north side. Each compartment connects to the corridor through a single \(80\,\text{cm}\) opening centered on its north wall, so the corridor and the two compartments form one continuous walkable region. The fully connected arena is treated as a single environment during both sensory-field generation and training. Therefore, the sensory field changes smoothly across the corridor. We train the network with \(800,000\) update steps, allowing grid-cell-like representations to stabilize while also tracking their drift over training.

\subsection{Three-dimensional traversal experiments}
\label{sup:3d}

For the 3D experiments, we use a cubic arena of size \(97 \times 97 \times 97\) bins, with a 5-voxel border giving a \(107^3\) ambient map. The four sensory populations use 3D smoothing widths \(\sigma \in \{8, 10, 12, 14\}\), each containing \(256\) cells. The boundary-aware diffusion process described in Section~\ref{sup:sensory} is applied with a volumetric \(3 \times 3 \times 3\) kernel. The motion input is 4-dimensional, consisting of a 3D unit movement direction and a scalar speed.

In the random-fly traversal experiment, the agent moves freely through the volume using the same smooth trajectory generator and boundary-avoidance scheme as in 2D. We use \(\mu_{\text{spd}} = 30\), \(\sigma_{\text{spd}} = 10\), and \(\alpha_v = 0.5\), and train the model for \(800{,}000\) steps with firing-rate noise \(\sigma_n = 0.8\). Under this traversal, place-like cells develop localized 3D firing fields, while grid-like cells develop locally ordered volumetric firing fields, consistent with the partially ordered 3D grid patterns reported in flying bats.


\subsection{Robustness of co-emergence}
\subsubsection{Region ratio}
\label{sup:region_ratio}

The default network uses \(N_{\text{sens-drv}} = N_{\text{rec-drv}} = 1024\). To examine whether this exact split affects co-emergence, we trained four additional models with different sensory-driven/recurrent-driven ratios:
\((N_{\text{sens-drv}}, N_{\text{rec-drv}}) \in \{(512,1536), (768,1280), (1280,768), (1536,512)\}\),
while holding the total number of hidden units fixed at \(N=2048\).

Across all four models, both place cells and grid cells emerged consistently. The corresponding numbers of classified place/grid cells were \(263/68\), \(302/43\), \(307/94\), and \(313/12\), respectively. Although our main experiments were conducted in the \(1024/1024\) setting for purposes of clarity, this experiment shows that co-emergence is not specific to that case. The reduced grid-cell count in the \(1536/512\) split suggests that having too few recurrent-driven units can limit grid-cell emergence, but the qualitative presence of both cell types remains consistent across the tested partitions.

\subsubsection{Disabling homeostatic synaptic scaling}
\label{sup:homeostasis}

Setting \(\eta = 0\) in Eq.~\ref{eq:homeostasis} removes the column-wise outgoing-budget rescaling. Over short training horizons (\(\le 50{,}000\) steps), grid- and place-like responses still emerge qualitatively. However, over longer training horizons, the recurrent dynamics gradually drift: units that initially exhibit hexagonally arranged firing fields lose contrast over time, with firing-field centers becoming progressively weaker until the periodic structure is no longer visually apparent. We also found that, without homeostatic scaling, switching to a new room can produce infinite losses, possibly because the network has overfit to the preceding room. Homeostatic scaling therefore acts as a soft stabilizer that preserves the co-emergent operating regime over extended training, rather than being strictly required for the initial formation of grid- or place-like responses.



\end{document}